\documentclass[sn-mathphys-num]{sn-jnl}

\usepackage{graphicx}%
\usepackage{multirow}%
\usepackage{amsmath,amssymb,amsfonts}%
\usepackage{amsthm}%
\usepackage{mathrsfs}%
\usepackage[title]{appendix}%
\usepackage{xcolor}%
\usepackage{textcomp}%
\usepackage{manyfoot}%
\usepackage{booktabs}%
\usepackage{algorithm}%
\usepackage{algorithmicx}%
\usepackage{algpseudocode}%
\usepackage{listings}%
\usepackage{natbib}
\usepackage{cancel}
\usepackage{bm}

\theoremstyle{thmstyleone}%
\theoremstyle{thmstyletwo}%

\theoremstyle{thmstylethree}%

\usepackage{amssymb}
\usepackage{amsmath}
\usepackage{tabularx}
\usepackage{bm}
\usepackage{hyperref}
\usepackage{ltxtable}
\usepackage{natbib}
\usepackage{longtable}
\usepackage[normalem]{ulem}
\usepackage{float}

\newcommand{\la}{\left\langle}
\newcommand{\ra}{\right\rangle}
\newcommand{\be}{\begin{equation}}
	\newcommand{\ee}{\end{equation}}
\newcommand{\bea}{\begin{eqnarray}}
	\newcommand{\eea}{\end{eqnarray}}
\newcommand{\ba}{\begin{array}}
	\newcommand{\ea}{\end{array}}
\newcommand{\bi}{\begin{itemize}}
	\newcommand{\ei}{\end{itemize}}
\newcommand{\ben}{\begin{enumerate}}
	\newcommand{\een}{\end{enumerate}}

\begin{document}

\title[Turbulence: A Nonequilibrium Field Theory]{Turbulence: A Nonequilibrium Field Theory}


\author*[1]{\fnm{Mahendra} \sur{Verma}}\email{mkv@iitk.ac.in}


\affil*[1]{\orgdiv{Department of Physics}, \orgname{Indian Institute of Technology Kanpur}, \orgaddress{\city{Kanpur}, \postcode{208016}, \state{UP}, \country{India}}}


\abstract{Tools of quantum and statistical field theories have been successfully ported to turbulence. Here, we review the key results of turbulence field theory.  \textit{Equilibrium field theory} describes thermalized spectrally-truncated Euler equation, where the equipartitioned Fourier modes generate zero energy flux. In contrast, \textit{nonequilibrium field theory} is employed for modelling of hydrodynamic turbulence (HDT), which has small viscosity. In HDT,  viscosity renormalization yields wavenumber-dependent viscosity and energy spectrum.  Field theory calculations also yields nonzero energy flux for HDT. These field theory computations have been generalized to other systems, e.g., passive scalar and magnetohydrodynamics. In this review, I cover these aspects, along with a brief coverage of weak turbulence and intermittency. }




\maketitle

\section{Introduction}\label{sec1}

Classical and quantum field theories are critical pillars of modern physics, and they successfully explain complex many-body systems in high-energy, condensed-matter, and statistical physics~\cite{Peskin:book:QFT,Goldenfeld:book,Fradkin:book:QFT}.  Notable results of field theory are the running coupling constants in quantum electrodynamics (QED) and quantum chromodynamics (QCD)~\cite{Peskin:book:QFT}; universality in second-order phase transition~\cite{Wilson:PR1974};  quantum Hall effect; and turbulence, a topic of the present review.   

Early applications of quantum field theories were for equilibrium systems, e.g., Ising Hamiltonian, free-electron theory, $\phi^4$ theory, and Hubbard model. Here,  the  fields are in equilibrium, which leads to the energy spectrum $E(k) \sim k^{d-1}$ in $d$ dimensions~\cite{Goldenfeld:book,Fradkin:book:QFT}.  Analogous equilibrium system in hydrodynamics is spectrally-truncated Euler equation~\cite{Cichowlas:PRL2005}.

Note, however, that many physical systems, including turbulence, are out of equilibrium.  Such systems are often forced and dissipative at different length scales, leading to uneven energy distribution across scales. For example, the energy spectrum for three-dimensional (3D) hydrodynamic turbulence (HDT) is $k^{-5/3}$, not equilibrium $k^{d-1}$.~\cite{Kolmogorov:DANS1941Structure,Kolmogorov:DANS1941Dissipation,Lesieur:book:Turbulence}  In addition, nonequilibrium systems often exhibit multiscale energy cascades, which are absent in equilibrium systems. In this review, we will contrast the equilibrium and nonequilibrium properties of turbulent flows.

Another common feature among field theory calculations is \textit{Perturbative expansion}~\cite{Peskin:book:QFT,Goldenfeld:book,Fradkin:book:QFT}. Here,  the contributions of the  nonlinear or interaction term  are computed perturbatively over the linear term.  The relative strength of the interaction term over the linear term is called  the \textit{coupling constant} of the theory. 
QED, conventional superconductors, weakly interacting Bose gases, and many other quantum systems have small coupling constants. Hence,  perturbation expansion is justified for such systems. In contrast, perturbative expansion is not employed to QCD because its  the coupling constant is larger than unity.  Interestingly, the coupling constant of HDT is of the order of unity. Yet,  perturbative turbulence field theory yields  reasonable results, which will be discussed in this review.   

Nonequilibrium  nature of turbulence makes its field theory interesting, as well as challenging with several unsolved problems. In this paper, I will review important developments in  field theory of turbulence. There are many books~\cite{Leslie:book,McComb:book:Turbulence,McComb:book:HIT,Sagaut:book}
 and review articles~\cite{Orszag:CP1973,Kraichnan:ROPP1980, Zhou:NASA1997,Zhou:PR2010,Zhou:PR2021,Eyink:RMP2006,Antonov:RNC2025} on this topic. Hence, for novelty, this review  attempts to bring in  new perspectives and presents a couple of new results. For example, I contrast turbulence field theory with equilibrium field theories,   and present  new and unpublished field-theoretic calculations on passive scalar and magnetohydrodynamic turbulence based on Craya-Herring basis.

After these introductory remarks, I introduce some  key  results of turbulence field theory, whose starting point is \textit{incompressible Navier-Stokes equations}: 
 {\bea
 		\frac{\partial {\bf u}}{\partial t} -  \nu \nabla^2 {\bf u} & = &   -\lambda {\bf u} \cdot \nabla {\bf u}  
 	- \nabla p + {\bf f},  \label{eq:NS}  \\
	\nabla \cdot {\bf u} & = & 0,
	\label{eq:del_u_0}
	\eea}	
where ${\bf u}, p$ are the velocity and pressure fields, respectively; $\lambda$ is the coupling constant;  $\nu$ is the kinematic viscosity; and ${\bf f}$ is the coloured noise with the following correlation:
\be
\la f_l({\bf x})  f_l({\bf x'}) \ra
\sim |{\bf x-x'}|^{y-d-z},
\label{eq:ff_corr_real}
\ee 
where $d$ is the space dimension; $y$ is the  exponent for the spatial force-force correlation;  and $z$ is the dynamic exponent ($\omega \sim k^z$, where $\omega$ is frequency). 

Equations~(\ref{eq:NS}, \ref{eq:del_u_0}) do not have a general mathematical solution. In fact, it is not yet known if Eqs.~(\ref{eq:NS}, \ref{eq:del_u_0}) admit a smooth solution for small viscosity~\cite{Frisch:book}. Fortunately, stalwarts like Taylor, Batchelor, Kolmogorov, Chandrasekhar, and others solved for the Green's and correlation functions under the assumption of homogeneity and isotropy. For brevity, we present only the Kolmogorov's K41 theory, according to which, for large-scale forcing with $\nu \rightarrow 0$, under the assumption of homogeneity and isotropy, the third-order structure function is~\cite{Kolmogorov:DANS1941Dissipation,Kolmogorov:DANS1941Structure,Frisch:book}
\be 
S_3(l) = \la | ({\bf u(x+l) - u(x)}) \cdot \hat{l} |^3 
\ra  = - \frac{4}{5} \Pi l,
\label{eq:S3_K41}
\ee
where $\hat{l}$ is unit vector along ${\bf l}$, and $\Pi$ is the energy flux in the inertial range.  Using Eq.~(\ref{eq:S3_K41}), it is easy to derive the energy spectrum for HDT as $k^{-5/3}$. We may call K41 theory as a field theory, with Eq.~(\ref{eq:S3_K41}) as an exact field-theoretic relation. However, this calculation differs from ones using standard perturbative method. Kraichnan pioneered turbulence field theory, starting with his influential paper on \textit{direct interaction approximation} (DIA).

 \citet{Kraichnan:JFM1959} employed DIA, a first-order perturbative field theory, to HDT and derived the equations for the Green's and correlations functions. The integral for the Green's function (\textit{self-energy integral}) diverges at small wavenumbers, which  called the \textit{infrared divergence} problem in field theory. To cure this problem,  \citet{Kraichnan:JFM1959}  introduced an infrared cutoff for the integral. But, this trick leads to $k^{-3/2}$ energy spectrum the contradicts observed $k^{-5/3}$ energy spectrum for HDT.   We will discuss Kraichnan's formalism in Section~\ref{sec:DIA}. Later,  Kraichnan performed more complex field theory computations---\textit{Mixed Lagrangian-Eulerian approach}~\cite{Kraichnan:PF1964Lagrangian_Eulerian}, \textit{Lagrangian-History Closure Approximation}~\cite{Kraichnan:PF1965Lagrangian_history}, \textit{Test Field Model}~\cite{Kraichnan:JFM1972}---to derive $k^{-5/3}$ energy spectrum for HDT.  Another important field theory work on turbulence is by   \citet{Wyld:AP1961} who  wrote Feynman diagrams to higher orders.  Unfortunately, Wyld did not close the equations to predict the Green's and correlation functions for HDT.

Renormalization groups (RG) provides a pathway to solve the infrared divergence problem. \citet{Yakhot:JSC1986} and \citet{Forster:PRA1977} applied RG  to hydrodynamic turbulence and derived interesting results. For example, \citet{Yakhot:JSC1986} obtained Kolmogorov's $k^{-5/3}$ energy spectrum 
 for $y=d$ [see Eq.~(\ref{eq:ff_corr_real})].  \citet{McComb:PRA1992Two_field, McComb:book:Turbulence}, and \citet{Zhou:PRA1988} employed recursive self-consistent RG  and showed that the renormalized viscosity scales as $k^{-4/3}$, and that the energy spectrum scales as $k^{-5/3}$. We will discuss these theories in Sections~\ref{sec:YO} and \ref{sec:RecursiveRG}. 
 
 \citet{Kraichnan:JFM1959}, \citet{Yakhot:JSC1986}, \citet{Forster:PRA1977}, \citet{McComb:PRA1992Two_field}, and  \citet{Zhou:PRA1988} employed perturbative expansion to the Navier-Stokes equation. This is in contrast to standard field-theory practice of using Hamiltonian, Lagrangian, or generating functionals. Note, however,  that  \citet{DeDominicis:PRA1979} and \citet{Martin:PRA1973} constructed  generating functionals for HDT using coloured noise or forcing  [Eq.~(\ref{eq:ff_corr_real})]. \citet{DeDominicis:PRA1979}  derived a relationship between the Green's function and correlation function using Callan-Symanzik equation, and Kolmogorov's spectrum for a special case of forcing. 
 
 Field-theoretic computations are quite tedious involving complex tensor algebra. Recently, \citet{Verma:PRE2024_d_dim} employed Craya-Herring basis that simplified the calculations tremendously. In this framework, pressure is eliminated automatically. More importantly, each  Craya-Herring component of the velocity field has its own renormalized viscosity.    I will cover these results in Section~\ref{sec:RG_CH}. Turbulence is a dynamic phenomena with nonzero multiscale energy transfers. Fortunately, turbulence energy transfers have been computed  using first-order perturbation, starting from \citet{Kraichnan:JFM1959}.  I will discuss these computations in Section~\ref{sec:ET}. In addition, I will discuss field theories of scalar and magnetohydrodynamic (MHD) turbulence in  Sections~\ref{sec:PS} and \ref{sec:MHD};  Craya-Herring basis simplifies these computations significantly and provide interesting relationships with HDT.
 
Now, I discuss some topics they are not detailed in this review.    Many real flows involve gravitational field, or magnetic field, or some other external fields that  make the flow anisotropic.  Anisotropic field theories are quite complex and they are beyond the scope of this review.  In addition, I do not delve into the first-principle derivation of multipoint correlation functions, a topic related to intermittency \cite{Lvov:PRL1996fusion,Falkovich:RMP2001}. In another front, weak turbulence theory employs field-theoretic tools with nonlinear term as a small perturbation over the linear term.  In Sections~\ref{sec:WT} and \ref{sec:Intermittency}, I will discuss  weak turbulence and intermittency very briefly. In Section~\ref{sec:other_FT}, I compare and contrast various field theories including turbulence.

The present review employs many symbols; I tabulate them in Table~\ref{tab:abbrev} for convenience of the reader. In the following section, I will introduce the basic equations used in this review.

\begin{table}[h]
	\caption{Abbreviation and symbols used in this review.}\label{tab:abbrev}%
	\begin{tabular}{@{}ll|ll@{}}
		\toprule
		Symbol & Stands for & Symbol & Stands for  \\
		\midrule
		CH & Craya-Herring & HDT & Hydrodynamic turbulence \\
		KE & Kinetic energy & ME & Magnetic energy \\
		RG & Renormalization group & NS & Navier-Stokes \\
		MHD & Magnetohydrodynamics & 		$c = k_{n+1}/k_n$ & wavenumber binning parameter \\
		\textbf{r}  & space coordinate  & \textbf{k} &wavenumber  \\
		\textbf{u}  &velocity field   & $u_1$ & CH basis: 1st component \\
		$d$ & space dimension   & $u_2$ &CH basis: 2nd component \\
		$\textbf{f}$ & force field to \textbf{u} & $\psi$ & Scalar field\\
		\textbf{b} & magnetic  field   & ${\bf z^\pm = u\pm b}$ & Els\"{a}sser variables\\
		$E(k)$ & 1D KE spectrum & $E({\bf k}) = \frac{1}{2} |{\bf u(k)}|^2$ & Modal KE  \\
		$|u_1({\bf k})|^2 = C_1({\bf k})$ &  Spectral correlation for $u_1({\bf k})$& $|u_2({\bf k})|^2=C_2(\textbf{k})$  & Spectral correlation for $u_2({\bf k})$ \\
		$C_1({\bf k}) = C_2({\bf k})$ & $C({\bf k})$ (isotropy) & 
		$G({\bf k}) $ & Green's function for \textbf{u} \\
		$G_1({\bf k}) $ & Green's function for $u_1$ & $G_2({\bf k}) $ & Green's function for $u_2$ \\
		$|\psi({\bf k})|^2 = C^\psi({\bf k})$ & Spectral correlation for $\psi({\bf k})$ & $G^\psi({\bf k}) $ & Green's function for $\psi$  \\
		$\bar{C}_{1,2}({\bf k},t-t')$ & two-time correlation functions for $u_{1,2} $ &
		$\bar{C}^\psi({\bf k},t-t') $ & two-time correlation function for $\psi$ \\
		$\bar{C}^\pm_{1,2}({\bf k},t-t')$ & two-time correlation functions for ${z}^\pm_{1,2} $ &
		$G^\pm_{1,2}({\bf k},t-t') $ & Green's function for ${z}^\pm_{1,2} $  \\
		$K_\mathrm{Ko}$ & Kolmogorov constant &$K_\psi$ &
		Obukhov-Corrsion constant \\
		$K^\pm$ & Kolmogorov constant for ${\bf z}^\pm $ && \\
		$\nu$ & Kinematic viscosity & $\nu_1$ & viscosity for $u_1$ \\
		 $\nu_2$ & viscosity for $u_2$ & $\nu_{1,2*}$ & RG constant for $\nu_{1,2}$ \\
		 $\kappa$ & Scalar diffusivity & $\kappa_*$ & RG constant for $\kappa$ \\
		 $\eta$ &  magnetic diffusivity & $\eta_\pm$ &  $(\nu \pm \eta)/2$ \\
		 $\eta_{1,2}$ & diffusivities for $({z^\pm})_{1,2}$ [$\la |z^+|^2 \ra  = \la |z^-|^2 \ra$]& $\eta_{1,2*}$ & RG constants for $\eta_{1,2}$ \\
		 $\Pi$ & Energy flux for \textbf{u} & $\Pi_\psi$ & Energy flux for $\psi$ \\
		 $\Pi_{u_1}$ & Energy flux of $u_1$ & 
		  $\Pi_{u_2}$ & Energy flux of $u_2$ \\
		  $\Pi_{z_1}$ & Energy flux of $z_1$ & 
		  $\Pi_{z_2}$ & Energy flux of $z_2$ ($E^+ = E^-$) \\
		 \botrule
	\end{tabular}
\end{table}

\section{Governing Equations}
\label{sec:eqns}

Field-theoretic methods often employ spatial and temporal averaging, which is sensible for homogeneous and isotropic systems.   In addition,  the system is assumed to be steady that leads to two-time temporal correlations and Green's functions being independent of absolute time, that is, $F(t,t') = F(t-t')$ for any function $F$.  This framework has been adopted in turbulence field theory. Most  field-theoretic works on turbulence are for incompressible hydrodynamics, which is described using incompressible Navier-Stokes (NS) equations [Eqs.~(\ref{eq:NS}, \ref{eq:del_u_0})].   In the following discussion, we present the equations employed in field-theoretic calculations.

\subsection{Incompressible Navier-Stokes Equation}

Field-theoretic calculations are often performed in Fourier space, where the equations for the incompressible Navier-Stokes equations are~\citep{Lesieur:book:Turbulence,Verma:book:ET}
\bea
(\partial_t +\nu k^2){\bf u} (\mathbf{k},t) 
& = & - i \lambda \int \frac{d\bf p}{(2\pi)^d}    \{ {\bf k} \cdot  {\bf u}({\bf q}, t) \} {\bf u}({\bf p}, t)    -i {\bf k} p (\mathbf{k},t)   + {\bf f}({\bf k},t) , \label{eq:uk}\\
{\bf k\cdot u} (\mathbf{k},t) & = & 0. \label{eq:k_uk_zero}
\eea 
Here, ${\bf k = p+q}$; $\lambda$ is the coupling constant; $d$ is the space dimension;  and $\nu$ is kinematic viscosity.  The transformations from real space to Fourier space and vice versa are as follows~\citep{Peskin:book:QFT}:
\bea
{\bf u}({\bf r},t) = \int \frac{d\bf k}{(2\pi)^d} {\bf u}({\bf k}, t) \exp( i {\bf k \cdot r});~~~
{\bf u}({\bf k}, t) =  \int d{\bf r} [{\bf u}({\bf r},t)   \exp( -i {\bf k \cdot r})],
\eea
and the pressure field is determined using the following equation~\cite{Lesieur:book:Turbulence}:
\be 
p(\mathbf{k},t) =   - \frac{\lambda}{k^2}  \int \frac{d\bf p}{(2\pi)^d}    \{ {\bf k} \cdot  {\bf u}({\bf q}, t) \}   \{ {\bf k} \cdot  {\bf u}({\bf p}, t) \}  ,
\label{eq:Pk}
\ee
with ${\bf k} \cdot {\bf f}({\bf k},t) =0$. \textit{Galilean invariance} of NS equation leads to $\lambda = 1$~\cite{Forster:PRA1977}; this result plays a critical role in turbulence field theory.

Many field-theoretic calculations have been performed in  frequency and wavenumber space, where the NS equations in tensorial form are~\cite{Kraichnan:JFM1959,Leslie:book}
\bea
	\left(-i\omega+\nu k^{2}\right)u_{l}(\hat{{k}}) 
	& = & - i \frac{\lambda}{2}  P_{lmn}(\mathbf{k})\int_{\hat{{p}}+\hat{{q}}=\hat{{k}}}d\hat{{p}}\left[u_{m}(\hat{{q}})u_{n}(\hat{{p}})\right] + f_l(\hat{{k}}) ,
	\label{eq:uk_omega} \\
	k_l u_{l}(\hat{{k}}) & = & 0,
\eea
where $\omega$ is the frequency,  $\hat{k} = ({\bf k}, \omega)$, $\hat{p}+\hat{q} = \hat{k}$, 
and
\begin{eqnarray}
	P_{lmn}(\mathbf{k}) & = & k_{m}P_{lm}(\mathbf{k})+k_{n}P_{lm}(\mathbf{k}),\label{eq:Pp} \\
	P_{lm}({\bf k}) & = & \delta_{lm} - \frac{k_l k_m}{k^2} , \label{eq:P_lm} \\
	d\hat{p} &= & \frac{1}{(2\pi)^{d+1}} d{\bf p}  d\omega_p.
\end{eqnarray}

Field-theoretic computations are quite complex. However,  Craya-Herring (CH) basis~\citep{Craya:thesis,Herring:PF1974,Sagaut:book}  simplifies these computations considerably~\cite{Verma:PRE2024_d_dim,Verma:arxiv2023}.  In 3D,  the CH basis vectors are:
\bea
\hat{e}_0({\bf k}) = \hat{k};~~
\hat{e}_1({\bf k}) = \frac{ \hat{k} \times \hat{n}}{|\hat{k} \times \hat{n}|};~~~
\hat{e}_2({\bf k})= \hat{e}_0({\bf k})  \times \hat{e}_1({\bf k}) ,   \label{eq:CH_basis_defn}
\eea 
where  the unit vector $\hat{k}$ is along  the wavenumber ${\bf k}$, and the unit vector $\hat{n}$ is chosen along any direction. An incompressible  flow has components $ u_1({\bf k}) $ and  $u_2({\bf k}) $ along the unit vectors $\hat{e}_1({\bf k})$ and $\hat{e}_2({\bf k})$, respectively.
\begin{figure}
	\begin{center}
		\includegraphics[scale = 1]{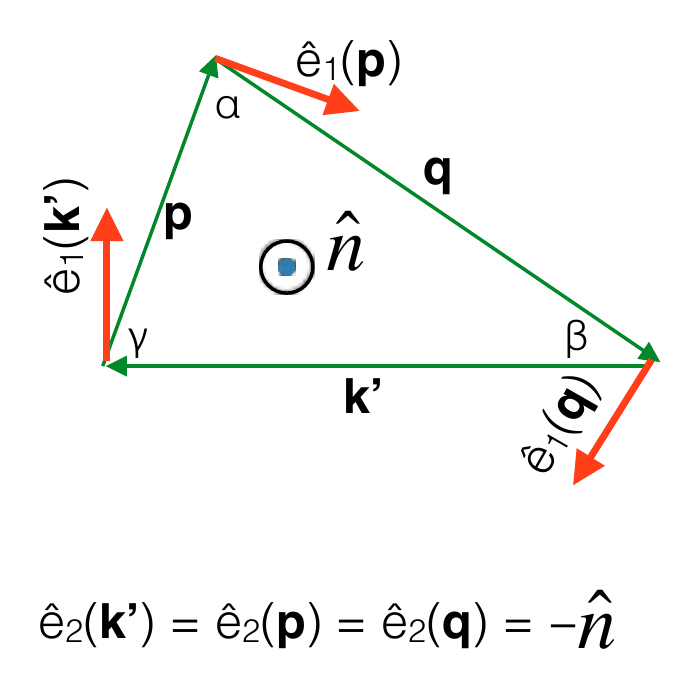}
	\end{center}
	\vspace*{0pt}
	\caption{Craya-Herring basis vectors  for the wavenumbers in an interacting wavenumber triad $({\bf k',p,q})$. From \citet{Verma:book:ET}. Reprinted with permission from Verma.}
	\label{fig:CH_triad}
\end{figure}
We consider an interacting wavenumber triad $({\bf k',p,q})$  with ${\bf k'+p+q} = 0$, and ${\bf k' = -k}$. We choose $\hat{n} =({\bf q \times p})/ |{\bf q \times p}|$~\citep{Waleffe:PF1992,Sagaut:book,Verma:book:ET}.
The Craya-Herring basis vectors for the interacting wavenumbers   are illustrated in Fig.~\ref{fig:CH_triad}.  Note that $ \alpha, \beta$,  $\gamma$ are the angles in front of $ k, p, q $ respectively.  \citet{Verma:book:ET} derived the following equations for the Craya-Herring components  $u_1$ and $u_2$ shown in Fig.~\ref{fig:CH_triad}:
\bea
(\partial_t +\nu k^2){u}_1({\bf k'},t) & = & i k' \sin(\beta-\gamma) u_1^*({\bf p},t)  u_1^*({\bf q},t) +f_1({\bf k'},t) ,
\label{eq:u1k_dot}  \\
(\partial_t +\nu k^2){u}_1({\bf p},t) & = &i p \sin(\gamma-\alpha) u_1^*({\bf q},t) u_1^*({\bf k'},t)  +f_1({\bf p},t)  , 
\label{eq:u1p_dot}\\
(\partial_t +\nu k^2){u}_1({\bf q},t) & = &i q \sin(\alpha - \beta)   u_1^*({\bf p},t) u_1^*({\bf k'},t) +f_1({\bf q},t) , 
\label{eq:u1q_dot}
\eea
\bea
(\partial_t +\nu k^2){u}_2({\bf k'},t) & = &i k'  \{ \sin \gamma u_1^*({\bf p},t)  u_2^*({\bf q},t) -\sin\beta u_1^*({\bf q},t)  u_2^*({\bf p},t)\} +f_2({\bf k'},t) , \label{eq:u2k_dot} 
 \nonumber \\ \\
(\partial_t +\nu k^2){u}_2({\bf p},t) & = &i p  \{ \sin \alpha u_1^*({\bf q},t)  u_2^*({\bf k'},t) -\sin\gamma u_1^*({\bf k'},t)  u_2^*({\bf q},t)\} + f_2({\bf p},t) , \label{eq:u2p_dot}  \nonumber \\ \\
(\partial_t +\nu k^2){u}_2({\bf q},t) & = &i q  \{ \sin \beta u_1^*({\bf k'},t)  u_2^*({\bf p},t) -\sin\alpha u_1^*({\bf p},t)  u_2^*({\bf k'},t)\} +f_2({\bf q},t) ,
\label{eq:u2q_dot}  \nonumber \\
\eea 
where $f_1$ and $f_2$ are force components along $\hat{e}_1$ and $\hat{e}_2$  respectively.  We need to integrate over all possible triads for the evolution of ${\bf u(k}, t)$.

For hydrodynamic turbulence, the \textit{modal energy}  for wavenumber ${\bf k}$ is 
\be 
E({\bf k}) = \frac{1}{2} |{\bf u(k)}|^2 =
\begin{cases}
	\frac{1}{2} |u_1({\bf k})|^2   & \text{for 2D} \\
	\frac{1}{2} |u_1({\bf k})|^2  + \frac{1}{2} |u_2({\bf k})|^2  & \text{for 3D}.
\end{cases}
\ee
For an isotropic  flow,  the spectral correlation $\la |u_i({\bf k})|^2 \ra $ for the CH components $u_i({\bf k})$ are equal\footnote{In this paper, we  denote $|u_1({\bf k})|^2 = C_1({\bf k})$ and $|u_2({\bf k})|^2 = C_2({\bf k})$.}. That is,
\be
\la |u_i({\bf k})|^2 \ra    \equiv  C({\bf k}) = C(k)
\ee
for all $i$'s. That is, $\la |u_i({\bf k})|^2 \ra  $ depends only on the magnitude of ${\bf k}$, which is denoted by $k$.  The total kinetic energy  of the flow is
\bea 
\frac{\la u^2 \ra}{2} = \int E(k) dk = \frac{1}{2} \int  \frac{d{\bf k }} {(2\pi)^{d}} (d-1) C({\bf k}) = \frac{1}{2} \frac{S_d}{(2\pi)^d} (d-1) \int dk  k^{d-1} C({\bf k}), 
\eea
where $E(k)$ is the one-dimensional (1D) shell spectrum, and $S_d = 2 \pi^{d/2}/\Gamma(d/2)$ is the surface area of the $d$-dimensional sphere.  The above equation yields the following  relationship between the modal energy and 1D energy spectrum~\citep{Kraichnan:JFM1959,Leslie:book,Verma:PR2004}:
\be
E(k) = \frac{(d-1)}{2}  \frac{S_d  k^{d-1} }{ (2\pi)^d} C({\bf k}).
\label{eq:Ek_Ck}
\ee
In Kolmogorov's theory of turbulence~\cite{Kolmogorov:DANS1941Dissipation,Kolmogorov:DANS1941Structure}, 1D energy spectrum is
\bea
E(k) & = & K_\mathrm{Ko}  \Pi^{2/3} k^{-5/3},
\label{eq:Ek_Kolm}
\eea
where $K_\mathrm{Ko}  $ is Kolmogorov's constant, and $\Pi$ is the inertial-range kinetic energy flux or the kinetic energy dissipation rate. Note, however, that the thermalized Euler turbulence (with $\nu=0$) admits equilibrium solution, for which 
\be 
E(k) \sim k^{d-1}.
\ee 
This equilibrium solution has zero energy flux~\cite{Cichowlas:PRL2005,Verma:arxiv2020_equilibrium}.

In Section~\ref{sec:DIA}, I present how \citet{Kraichnan:JFM1959}  computed the Green's function and renormalized viscosity using Direct Interaction Approximation (DIA).

\section{Direct Interaction Approximation (DIA)}
\label{sec:DIA}

The presentation in this section is a minor modification of \citet{Kraichnan:JFM1959}'s original calculation. \citet{Kraichnan:JFM1959} started with Eq.~(\ref{eq:uk_omega}), and $\lambda = 1$ due to Galilean invariance.  The \textit{bare Green's function} for the linearized NS equation is 
\be
G(\hat{k}) = \frac{1}{-i\omega+\nu k^{2}}.
\ee 
\citet{Kraichnan:JFM1959} treated the nonlinear term of the NS equation as perturbation.  In addition, the velocity field is assumed to be homogeneous and quasi-Gaussian, which leads to
\be 
\la u_{m}(\hat{{q}})u_{n}(\hat{{p}}) \ra  = C(\hat{q}) P_{mn}({\bf q}) \delta({\bf p+q}) \delta({\omega_p+\omega_q}),
\label{eq:DIA_corr}
\ee
where $\hat{p} = (\omega_p, {\bf p})$ and  $\hat{q} = (\omega_q, {\bf q})$. Under quasi-Gaussian approximation, the third-order correlation $ \la u_{m}(\hat{{q}})u_{n}(\hat{{p}}) u_{s}(\hat{{p}}) \ra$ is nonzero, and it is computed by expanding the velocity field perturbatively, and then reducing the fourth-order correlations to a sum of  products of two second-order correlations.

\begin{figure}
	\begin{center}
		\includegraphics[scale = 0.5]{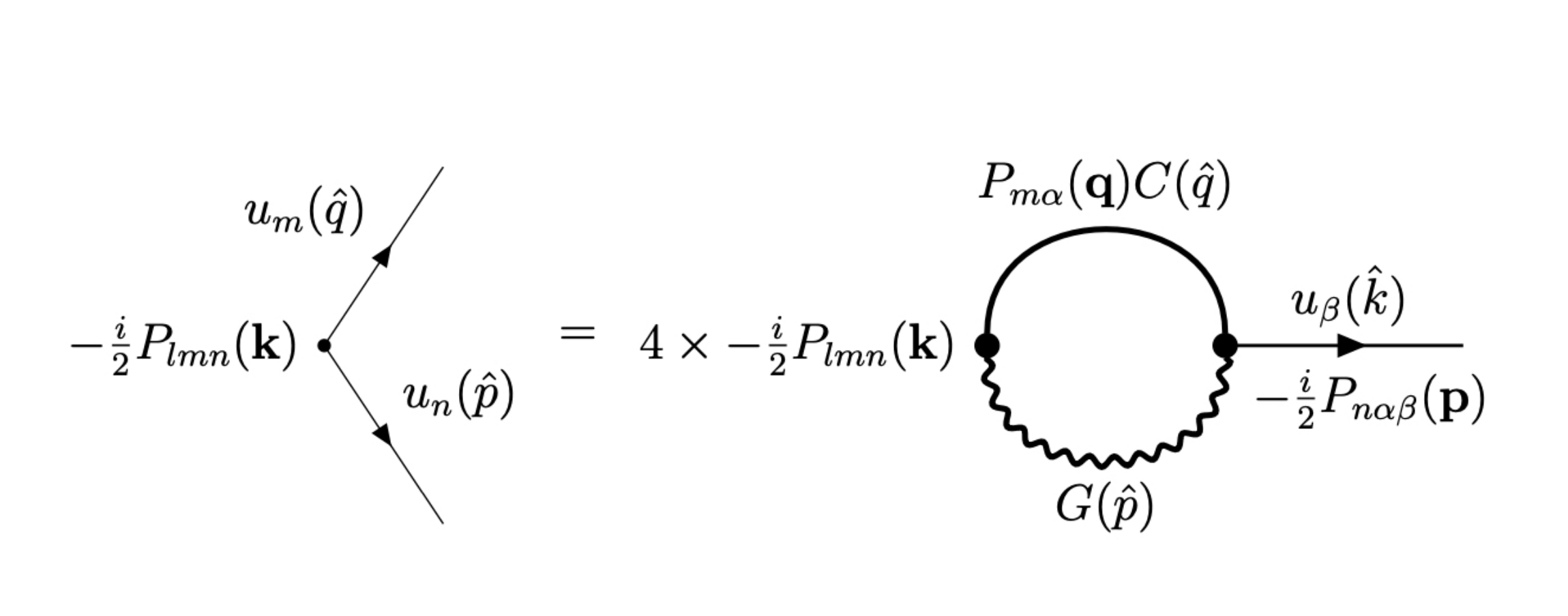}
	\end{center}
	\vspace*{0pt}
	\caption{Feynman diagrams associated with  the computation of effective viscosity and Green's function  using direct interaction approximation (DIA). }
	\label{fig:DIA_feyn}
\end{figure}

To zeroth order, the ensemble-averaged value of the nonlinear term of Eq.~(\ref{eq:uk_omega}) vanishes, i.e.,
\be
P_{lmn}(\mathbf{k}) \la u_{m}(\hat{{q}})u_{n}(\hat{{p}}) \ra =0
\label{eq:zeroth_order_DIA}
\ee
because  of Eq.~(\ref{eq:DIA_corr}) and $P_{lmn}(\mathbf{k} = 0)= 0$~\cite{Kraichnan:JFM1959,Leslie:book,Forster:PRA1977}.  Hence, \citet{Kraichnan:JFM1959} went to the next order, which is represented using the  Feynman diagram of Fig.~\ref{fig:DIA_feyn}.
To the first order, the transformed equation for the velocity field is~\cite{Kraichnan:JFM1959}
\bea
\left(-i\omega+\nu k^{2}\right) u_{l}(\hat{{k}}) 
& = & - u_{\beta}(\hat{{k}})   \int d\hat{{q}} \left[G(\hat{{p}})C(\hat{{q}}) A_{l \beta}({\bf k,p,q})   \right]  + f_{l}(\hat{{k}}) ,
\label{eq:dnu}
\eea
where
\bea
 A_{l \beta}({\bf k,p,q}) & = & P_{lmn}(\mathbf{k}) P_{n\alpha \beta }(\mathbf{p})  P_{m \alpha}(\mathbf{q})   
 \eea
 with the \textit{effective} or \textit{dressed Green's function} as
 \be
G(\hat{k}) = \frac{1}{-i\omega+\nu(k) k^{2}},
\ee 
and \textit{dressed correlation function} as
\bea
\la u_i^*(\hat{k}) u_j(\hat{k}) \ra & = & \frac{1}{-i\omega + \nu(k)k^2 } {C}({\bf k}) P_{ij}({\bf k}) .
\label{eq:C_q} 
\eea
In terms of  time variable,
\bea
G({\bf k}, t-t') & = & H(t-t') \exp[-\nu(k) k^2 (t-t') ],  \label{eq:Gk_tt'} \\
\bar{C}({\bf k}, t-t') & = & C({\bf k}) \exp[-\nu(k) k^2 (t-t')],  \label{eq:Ck_tt'}
\eea
where $H(t-t')$ is the heaviside function, which is 0 for $t<t'$ and 1 for $t>t'$. The above equations show that the two-time Green's function and correlations function decay with a time scale of $(\nu(k) k^2)^{-1}$, where $\nu(k)$ is the effective or dressed viscosity. This  important assumption, employed in almost all turbulence calculations~\cite{Kraichnan:JFM1959,Leslie:book,Lesieur:book:Turbulence}, is an extrapolation of \textit{fluctuation-dissipation theorem}  to systems far from equilibrium~(see Section \ref{sec:other_FT}).

The nonlinear term of Eq.~(\ref{eq:dnu}) is proportional to the velocity field. However, the velocity components in the left-hand side and right-hand side of Eq.~(\ref{eq:dnu})  are unequal ($u_l$ and $u_\beta$, respectively). Therefore, the viscosity correction   due to the nonlinear term is a second-rank tensor:
\be
\nu_{ l \beta}({\bf k}) =  \int d\hat{{q}}  G(\hat{{p}})C(\hat{{q}}) A_{l \beta}({\bf k,p,q})  .
\ee
This feature appear surprising for an isotropic flow, but this is reasonable because the renormalized parameter in the plane of the triad $({\bf k,p,q})$ may differ from that perpendicular to this plane\footnote{Recently,~\citet{Verma:PRE2024_d_dim} observed such behaviour in his calculations based on Craya-Herring basis. See Section~\ref{sec:RG_CH}.}. However,~\citet{Kraichnan:JFM1959} and many other researchers have assumed the \textit{dressed viscosity} to be an isotropic tensor, i.e.,
\be 
\nu_{ l \beta}(\hat{k})  = \nu(\hat{k}) P_{l \beta}({\bf k}),
\label{eq:nu_tensor}
\ee
with
\be
\nu(\hat{k})  =  \frac{1}{(d-1)k^2 }\int d\hat{{p}} Q(k,p,q) G(\hat{{p}})C(\hat{{q}}) ,
\label{eq:nu_isotropic}
\ee
where
\bea
Q({ k,p,q}) & = & P_{lmn}(\mathbf{k}) P_{n\alpha \beta }(\mathbf{p})  P_{m \alpha}(\mathbf{q})   P_{l \beta}({\bf k})
=  pk [(d-3)z + 2 z^3 + (d-1) xy] .
\label{eq:Q_DIA}
\eea

Using Eqs.~(\ref{eq:dnu}, \ref{eq:nu_tensor}, \ref{eq:nu_isotropic}), Eq.~(\ref{eq:uk_omega}) is rewritten in terms of dressed viscosity as
\bea
\left(-i\omega+\nu k^{2} + \nu(k) k^2 \right) u_{l}(\hat{{k}}) 
& = &   f_{l}(\hat{{k}}) 
\eea 
with
\bea
\nu(k) & = &   \frac{1}{(d-1)k^2 } \int \frac{d {\bf q} }{(2\pi)^d}
\frac{ Q(k,p,q) C(q)}{\nu(p) p^2 + \nu(q) q^2}
\label{eq:nu(k)_DIA}
\eea
under the long time limit ($\omega \rightarrow 0$) \cite{Kraichnan:JFM1959,Leslie:book,Orszag:CP1973,Verma:arxiv2005}. \citet{Kraichnan:JFM1959}  attempted the  self-consistent solution to Eq.~(\ref{eq:nu(k)_DIA}) using $C(k) $ of Eq.~(\ref{eq:Ek_Ck}, \ref{eq:Ek_Kolm}) and the following formula for $\nu(k)$  (based on Kolmogorov's spectrum):
\bea
\nu(k) & = & \nu_* \sqrt{K_\mathrm{Ko}} \Pi_u^{2/3} k^{-4/3}  ,
\eea
where  $\nu_*$ is a nondimensional constant. Further, he  nondimensionalized Eq.~(\ref{eq:nu(k)_DIA}) using $p = kp'$ and $q=k q'$ that yields
\bea
\nu_*^2 & = &  \frac{2 S_{d-1}}{S_d (d-1)^2} 
\int_0^\infty  dq' q'^{d-1} \int_{-1}^1 dy (1-y^2)^{(d-3)/2}
\frac{ Q({1,p',q'}) q'^{-2/3-d} }{ p'^{2/3} +  q'^{2/3}}, 
\label{eq:nu_DIA} 
\eea
where $y = \cos\beta$ with $\beta$ being the angle between ${\bf k}$ and ${\bf q}$ (see Fig.~\ref{fig:CH_triad}).
 
Unfortunately, the integral of Eq.~(\ref{eq:nu_DIA}) has infrared divergence near $p'=0$. \citet{Kraichnan:JFM1959} tried to cure this divergence  by introducing a lower or infrared cutoff for $p'$, but  this trick leads to $k^{-3/2}$ energy spectrum  that contradicts the popular Kolmogorov's theory.  \citet{Leslie:book} attempted a $k$-dependent lower cutoff for $p'$, i.e., $p' \ge ck$ ($c$ is a constant) that leads to Kolmogorov's spectrum. Inspired by  the infinity cancellations  employed in quantum electrodynamics~\cite{Peskin:book:QFT},
\citet{Leslie:book} suggested splitting the integral of Eq.~(\ref{eq:nu_DIA}) into singular and nonsingular parts;  unfortunately, this idea has not been implemented  till date.

Renormalization-Group (RG) is employed to solve the above  divergence problem~\cite{Peskin:book:QFT}. In Section~\ref{sec:RG}, I describe viscosity renormalization  based on Wilson's wavenumber RG scheme.

\section{Renormalization Group Analysis of Hydrodynamic Turbulence}
\label{sec:RG}

\begin{figure}
	\begin{center}
		\includegraphics[scale = 1]{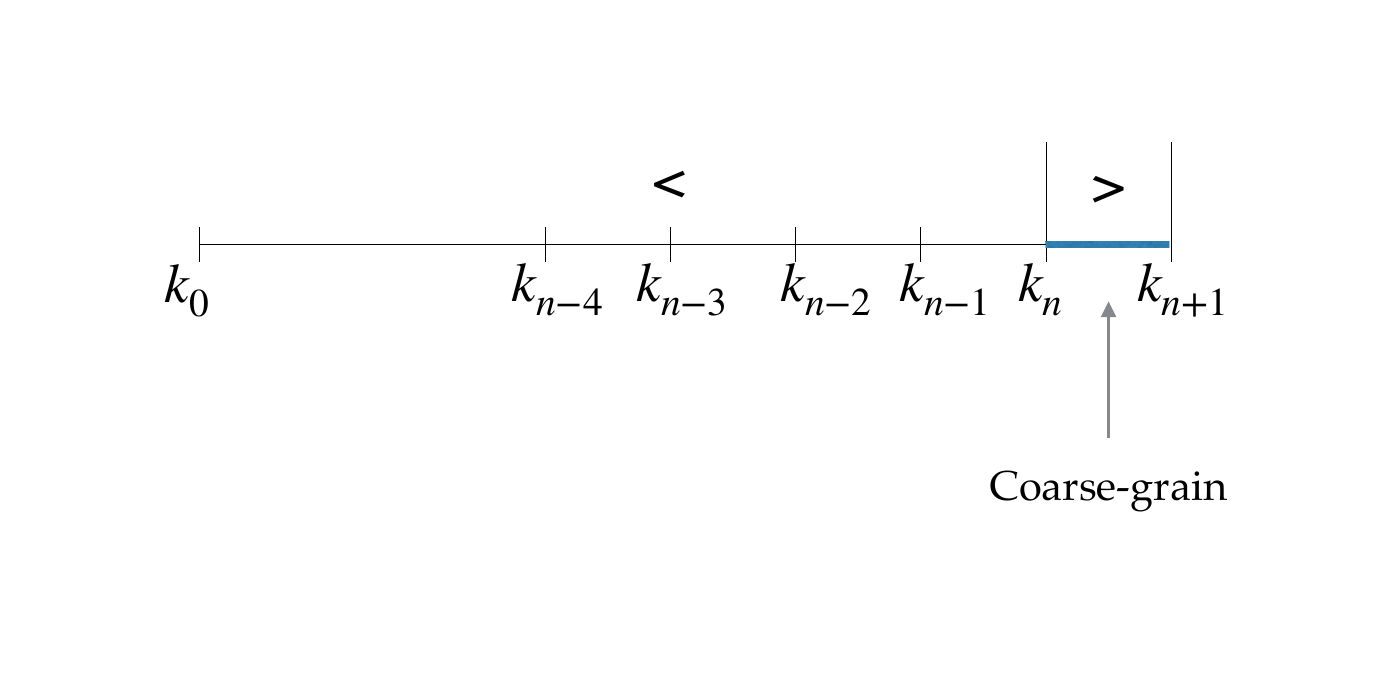}
	\end{center}
	\vspace*{0pt}
	\caption{ In wavenumber renormalization, the modes in the wavenumber band $(k_n, k_{n+1})$, denoted by $>$, are coarse-grained. The coarse-graining operation leads to enhancement of effective viscosity for wavenumbers $k < k_n$, denoted by $<$.  From \citet{Verma:PRE2024_d_dim}. Reprinted with permission from APS. }
	\label{fig:wavenumber_shells}
\end{figure}

In this section, I will cover prominent HDT RG schemes, mainly that of Yakhot and Orszag (YO)~\cite{Yakhot:JSC1986}, \citet{McComb:book:Turbulence}, \citet{Zhou:PRA1988}, \citet{DeDominicis:PRA1979}, and  \citet{Verma:PRE2024}.
Following Wilson's wavenumber renormalization, the wavenumber space is  logarithmic-binned  with $k_m = k_0 c^m$ and $c > 1$. For the wavenumber  up to $k_{n+1}$ shown in Fig.~\ref{fig:wavenumber_shells}, Eq.~(\ref{eq:uk_omega}) is rewritten as
\bea
\left(-i\omega+\nu^{(n+1)} k^{2}\right)u^<_{l}(\hat{{k}}) 
& = & f_l(\hat{{k}}) - i \frac{\lambda}{2}  P_{lmn}(\mathbf{k})\int_{\hat{{p}}+\hat{{q}} + \hat{{k}}}d\hat{{p}}\left[u^<_{m}(\hat{{q}})u^<_{n}(\hat{{p}})
+ u^>_{m}(\hat{{q}})u^>_{n}(\hat{{p}}) \right.   \nonumber \\
&& \hspace{1in} + \left. u^<_{m}(\hat{{q}})u^>_{n}(\hat{{p}})
+ u^>_{m}(\hat{{q}})u^<_{n}(\hat{{p}}) \right]  
\label{eq:RG1}
\eea
with $ \hat{q} = \hat{k} - \hat{p}$.  The convolutions in the  above equation  involves  four sums with wavenumbers \textbf{p} and \textbf{q} belonging to $<$ or $>$ regions of Fig.~\ref{fig:wavenumber_shells}. Now, we ensemble-average the fluctuations in wavenumber band $(k_n, k_{n+1})$, after which $\nu^{(n)}_1$ is the viscosity for the wavenumbers $(k_0,k_{n})$.  The above averaging process is called \textit{coarse graining}.

For the coarse-graining process,  it is assumed that  $u^{>}({\bf  k}, t) $ is  time-stationary, homogeneous, isotropic, and Gaussian with zero mean, and that $u^<({\bf  k}, t)$  is unaffected by coarse-graining~\citep{Wilson:PR1974,McComb:book:Turbulence,Zhou:PR2010}.   Another assumption is that the   correlation between $<$ and $>$ modes is weak.  Hence,
\bea
\la u_l^{>}({\bf  k}, t) \ra & = &0, \\
\la u_l^{<}({\bf  k}, t) \ra & = & u_1^{<}({\bf  k}, t), \\
\la u_l^{*<}({\bf  p}, t)  u_m^{*<}({\bf q}, t) \ra & = & u_l^{*<}({\bf  p}, t)  u_m^{*<}({\bf q}, t), \\
\la u_l^{*<}({\bf  p}, t)  u_m^{*>}({\bf q}, t) \ra & = &
u_l^{*<}({\bf  p}, t)  \la u_m^{*>}({\bf q}, t) \ra = 0, \\
\la u_l^{*>}({\bf  p}, t)  u_m^{*<}({\bf q}, t) \ra & = &
\la u_l^{*>}({\bf  p}, t)  \ra u_m^{*<}({\bf q}, t) = 0 .
\eea
Substitution of the above relations in Eq.~(\ref{eq:RG1})  yields
\bea
\left(-i\omega+\nu^{(n)} k^{2}\right)u^<_{l}(\hat{{k}}) 
& = & f_l(\hat{{k}}) - i \frac{\lambda}{2}  P_{lmn}(\mathbf{k})\int_{\hat{{p}}+\hat{{q}} + \hat{{k}}}d\hat{{p}}\left[u^<_{m}(\hat{{q}})u^<_{n}(\hat{{p}}) \right],
\label{eq:RG2}
\eea
where 
\be
\nu^{(n)} =   \nu^{(n+1)} + i  \frac{\lambda}{2 k^2}  P_{lmn}(\mathbf{k})\int^{\Delta}d\hat{{p}}\la u^>_{m}(\hat{{q}})u^>_{n}(\hat{{p}}) \ra =  \nu^{(n+1)} + \delta \nu^{(n)} ,
\label{eq:nu_k_second_integral}
\ee
with the integral performed over the coarse-grained region ($\Delta$). This integral provides the viscosity correction ($\delta \nu^{(n)}$).

At this point, we pause our discussion on HDT and go to Euler turbulence. The velocity field of thermalized spectrally-truncated Euler equation is  Gaussian. Hence, $\la u^>_{m}(\hat{{q}})u^>_{n}(\hat{{p}}) \ra = 0$ leading to $\delta \nu^{(n)} =0$. Thus, the viscosity of Euler equation remains unchanged at $\nu =0$; this is the result of \textit{equilibrium field theory}.

Under the quasi-Gaussian approximation, the  integral of Eq.~(\ref{eq:nu_k_second_integral})  vanishes to the zeroth order. Hence, the integral  is expanded to  the first-order. Researchers have adopted various tactics to compute $\delta \nu^{(n)}$. In the following subsections, we will describe the  schemes adopted by \citet{Yakhot:JSC1986},  \citet{McComb:PRA1983}, \citet{Zhou:PRA1988}, \citet{Verma:PRE2024_d_dim}, \citet{Martin:PRA1973}, and \citet{DeDominicis:PRA1979}. We start with \citet{Yakhot:JSC1986}'s RG scheme. 

\subsection{RG Scheme of Yakhot and Orszag}
\label{sec:YO}

\citet{Yakhot:JSC1986} (YO) employed dynamical RG framework~\cite{Hohenberg:RMP1977} that includes renormalization of viscosity, vertex, and forcing amplitude. In the NS equation, the vertex correction is absent due to the Galilean invariance. Therefore, \citet{Yakhot:JSC1986} computed the corrections to  viscosity  and forcing amplitude.

Following statistical field theory, \citet{Yakhot:JSC1986} employed  the following random force:
\be
\la f_l({\bf k}, \omega)  f_m({\bf k'}, \omega') \ra = 
2D k^{-y} P_{lm}({\bf k})  \delta({\bf k + k'}) \delta(\omega + \omega').
\ee
 Equation~(\ref{eq:ff_corr_real}) provides the real space correlation for the above force (after the frequency integral). The velocity field in terms of  Green's function is
\be
u_l(\hat{k}) = G(\hat{k}) f_l(\hat{k}),~~~\mathrm{or}~~~
u_l({\bf k}, \omega) = G({\bf k}, \omega) f_l ({\bf k},  \omega).
\ee
\citet{Yakhot:JSC1986} employed RG that includes (a) parameter corrections under coarse-graining, and then (b) system rescaling to get back to the original size.  These steps are  same as those in Wilson's framework~\cite{Wilson:PR1974}. We brief these two steps in the following discussion.

\subsubsection{Coarse-graining}

\begin{figure}
	\begin{center}
		\includegraphics[scale = 0.8]{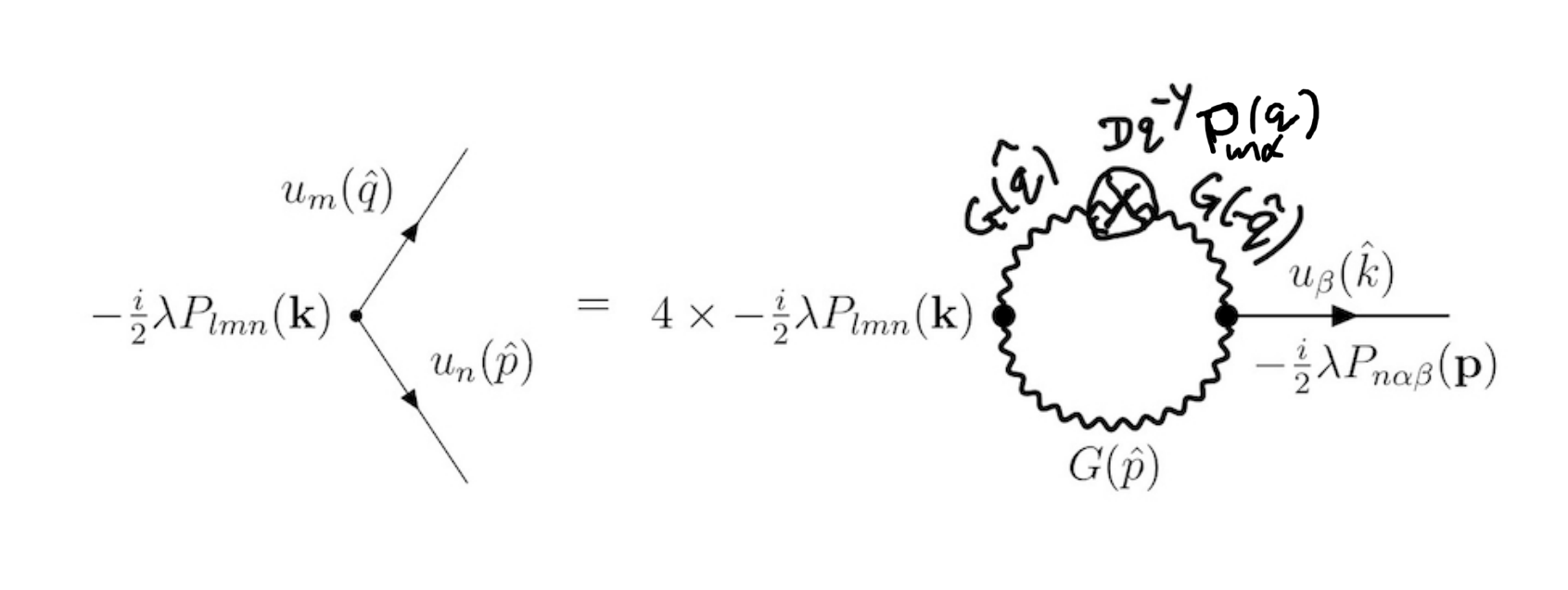}
	\end{center}
	\vspace*{0pt}
	\caption{Feynman diagrams associated with  the computation of renormalized viscosity by \citet{Yakhot:JSC1986}. }
	\label{fig:YO_feyn}
\end{figure}

In Eq.~(\ref{eq:nu_k_second_integral}), to zeroth order $\delta \nu^{(n)} =0$ because of the same reasons given in Section~\ref{sec:DIA} [refer to Eq.~(\ref{eq:zeroth_order_DIA})].  Therefore, \citet{Yakhot:JSC1986} computed the viscosity correction to the next order, which is represented by the Feynman diagram of Fig.~\ref{fig:YO_feyn}.  Here we assume \textit{large time limit}, which is
\be 
 \nu(k) = \lim_{\omega \rightarrow 0} 
\nu(k,\omega) .
\label{eq:nu_omega_0}
\ee 
Under these assumptions, the formula for the viscosity correction is
\bea
\delta \nu(k) & = & \lambda^2 \int \frac{d{\bf q}d \omega_q}{(2\pi)^{d+1}} Q({ k,p,q}) 
G(\hat{p}) G(\hat{q}) G(-\hat{q}) 
2D q^{-y} \nonumber \\
& = & \frac{\lambda^2}{2 \nu^2} \int d{\bf q}
Q({k,p,q}) D \frac{q^{-y-2}}{p^2+q^2}  \nonumber  \\
& = & \frac{\lambda^2}{2 \nu^2} S_d \int_{\Lambda}^{\Lambda c} q^{d-1} dq  
\int_{-1}^1 dz 
(1-z^2)^{(d-3)/2} Q({k,p,q}) D \frac{q^{-y-2}}{p^2+q^2},
\nonumber \\
&= & \nu \bar{\lambda}^2 A_d \frac{e^{\epsilon l} -1}{\epsilon},
\label{eq:d_nu_YO}
\eea
where $c = e^l$, $z = \cos \gamma$ (see Fig.~\ref{fig:CH_triad}), $Q(k,p,q)$ are given in Eq.~(\ref{eq:Q_DIA}), and   
\bea
\epsilon  & = &  4+y-d, 
\label{eq:YO_eps} \\
 \bar{\lambda}^2 & = & \frac{\lambda^2 D}{\nu^3 \Lambda^\epsilon} ,  \label{eq:lambda_bar} \\
 A_d & = & \frac{1}{2} \frac{d^2-d-\epsilon}{d(d+2)} \frac{S_d}{(2\pi)^{d}}.
\eea
Note that $ \bar{\lambda}$ is the nondimensional coupling constant for HDT.

In Eq.~(\ref{eq:nu_k_second_integral}), Yakhot and Orszag replaced $\nu^{(n+1)}$  with $\nu$, and set $k_{n} \rightarrow \Lambda$ and $k_{n+1} \rightarrow \Lambda c$. Here, we have skipped some factors in Eq.~(\ref{eq:d_nu_YO}), which have been absorbed in $A_d$. With $\delta \nu(k)$, the revised viscosity $\nu^<$ [$\nu^{(n)}$ of Eq.~(\ref{eq:nu_k_second_integral})] is
\be
\nu^< = \nu \left[ 1+\bar{\lambda}^2 A_d \frac{e^{\epsilon l} -1}{\epsilon} \right] . \label{eq:nu<}
\ee
The other two parameters to be renormalized are $\lambda$ and $D$. Galilean invariance leads to constancy of $\lambda$, which is
\be 
\lambda^< = \lambda.  \label{eq:lambda<}
\ee
\citet{Yakhot:JSC1986} argued that $D$ too remains unrenormalized, or
\be 
D^< = D. \label{eq:D<}
\ee
Equation~(\ref{eq:D<}) is consistent with Kolmogorov's theory of turbulence where forcing, applied at large scales, is absent in the inertial range.  Hence, the forcing amplitude is not renormalized.

\subsubsection{Rescaling}

Under coarse-graining, the wavenumber range $(k_0, k_{n+1})$ shrinks to $(k_{0}, k_{n})$ (by a factor $c^{-1}$). Following Wilson~\cite{Wilson:PR1974}, to rescale the system to its original size, we perform $k \rightarrow k' = k c$ that corresponds to $x = x' c$. In all, we rescale $x, t, u$, and $f$ as follows:
\be
x = x' c;~~t = t'c^z;~~u = u' c^\chi;~~f = c^{\frac{y-d-z}{2}} f' ,
\label{eq:rescale}
	\ee
where $z$ and $ \chi$ are constants that are determined using the RG analysis.  Note that $f$ scaling follows from Eq.~(\ref{eq:ff_corr_real}). Therefore, under rescaling, the NS equation transforms to
\bea
	\frac{\partial {\bf u'}}{\partial t'} - \nu^< c^{z-2}  \nabla^2 {\bf u}' & = &   -\lambda^< c^{\chi+z-1} [ {\bf u'} \cdot \nabla' {\bf u'}  
- \nabla'  p'] + c^{\frac{z+y-d}{2} - \chi } f'.
\label{eq:NS_rescaled}
\eea
Hence, a combination of coarse-graining and rescaling yields~[Eqs.~(\ref{eq:nu<}, \ref{eq:lambda<}, \ref{eq:D<}, \ref{eq:NS_rescaled})]
\bea
\nu' & = & \nu^< c^{z-2}, \\
\lambda'  & =  & \lambda^< c^{\chi+z-1}, \\
D' & = & D^<  c^{z+y-d- 2\chi}.
\eea
Using  $c = e^l$ and taking the limit $l \rightarrow 0$, we obtain
\bea
\frac{d \nu}{dl} & = & \nu \left[z-2 + A_d \bar{\lambda}^2\right] , \label{eq:dnu_dl} \\
\frac{d \lambda}{dl} & = & \lambda \left[\chi + z-1\right],
\label{eq:dlambda_dl} \\
\frac{d D}{dl} & = & D \left[ z+y-d -2 \chi \right] .
\label{eq:dD_dl}
\eea

Equations~(\ref{eq:dnu_dl}, \ref{eq:dlambda_dl}, \ref{eq:dD_dl}) have a trivial fixed point,  ($\nu=0, \lambda = 0, D =0$), and the following nontrivial fixed point:
\bea
z & = & 2 - \frac{\epsilon}{3} , 
\label{eq:z_unstable} \\
\chi & = &  \frac{\epsilon}{3} -1,
\label{eq:chi_unstable} 
\eea
where $\epsilon = 4+y-d$ [see Eq.~(\ref{eq:YO_eps})].  We take derivative of $\bar{\lambda}$ [Eq.~(\ref{eq:lambda_bar})] near the unstable fixed point [Eqs.~(\ref{eq:z_unstable}, \ref{eq:chi_unstable})] that yields  
\be 
\frac{d\bar{\lambda}}{dl} = \frac{\bar{\lambda}}{2} (\epsilon - 3 A_d \bar{\lambda}^2).
\label{eq:dlambda_bar_dl}
\ee
Hence, for $\epsilon <0$, $\bar{\lambda} \rightarrow 0$  (trivial  fixed point: $\nu=0, \lambda = 0, D =0$). Therefore, $\epsilon <0$ leads to decoupled thermalized Fourier modes, similar to the Gaussian fixed point of $\phi^4$ theory~\cite{Wilson:PR1974,Peskin:book:QFT,Goldenfeld:book}.  However, when $\epsilon >0$, $\bar{\lambda}$ moves to
\be 
\bar{\lambda} =\sqrt{\frac{\epsilon}{3 A_d}},
\label{eq:lambda_unstable}
\ee 
thus making the nonlinear term relevant.  The above calculation yields the RG fixed point for the nonequilibrium solution [Eqs.~(\ref{eq:z_unstable}, \ref{eq:chi_unstable}, \ref{eq:lambda_unstable})], similar to $\phi^4$ theory~\cite{Wilson:PR1974,Peskin:book:QFT,Goldenfeld:book}. Note that  $\bar{\lambda} = O(1)$ for the unstable fixed point.

Using this unstable RG fixed point, we derive the energy spectrum for HDT as follow. Using Eq.~(\ref{eq:rescale}) we derive 
\be 
\frac{u}{u'} = c^\chi = \left( \frac{k}{k'} \right)^{-\chi} \implies u_k \propto k^{-\chi} .
\ee
Hence,
\bea
E(k) & \sim & u_k^2/k \sim k^{-2\chi-1} \sim k^{-2 \epsilon/3 +1} .
\eea
We recover Kolmogorov's $k^{-5/3}$ spectrum when $\epsilon = 4$,  or $y=d$. 
For this case,  the forcing spectrum is $k^{-d}$, which is dominantly at large scales, as in Kolmogorov's theory of turbulence.  Hence, YO's RG results are consistent with Kolmogorov's theory of turbulence.  

In addition, for the Kolmogorov's spectrum, 
\be 
z = 2 - \frac{\epsilon}{3} = \frac{2}{3}~~~\implies 
\omega = k^z = k^{2/3},
\ee
leading to the dynamic exponent to be 2/3. Also, the renormalized viscosity (without rescaling) is
\bea
\nu^< = \nu(k) & \sim & \nu' b^{2-z} \sim k^{-4/3}
\eea
because $\nu'$ is a constant.

The external force injects energy   to the mode \textbf{u}(\textbf{k}) with a rate of $\la |{\bf f(k)}|^2 \ra$. Here, we focus on the special case $y=d$ that yields $k^{-5/3}$ spectrum. The energy injection rate to the flow in a wavenumber shell $(k_0,k)$ is~\cite{Yakhot:JSC1986}
\bea
\epsilon_\mathrm{inj}(k) & = & 2 D_0 \int_{k_0}^k k'^{-y} d{\bf k'} \nonumber \\
& = & 2 S_d D_0 \int_{k_0}^k k'^{-y+d-1} dk'
\nonumber \\
& = & 2 S_d D_0 \ln \frac{k}{k_0},
\label{eq:Dk_YO}
\eea
which is a slowly-varying logarithmic function of $k$, and it can be assumed to a constant.  

According to the variable energy flux framework~\cite{Verma:JPA2022},
\be 
\frac{d \Pi(k)}{dk} = \mathcal{F}(k) - D(k)
\label{eq:VEF}
\ee
where $\mathcal{F}(k)$ and $D(k)$ are, respectively, the energy injection rate and  dissipation rate to the wavenumber shell  of radius $k$. In YO formalism with $y=d$, we can assume that $\mathcal{F}(k) \rightarrow 0$ (due to steepness of $\mathcal{F}(k)$) and $D(k) \rightarrow 0$  in the inertial range. Hence, we obtain a nearly constant inertial-range energy flux:
\be 
\Pi(k)  \approx \epsilon_\mathrm{inj}.
\ee
Thus, YO formalism yields similar results as Kolmogorov's theory. Note, however, that in the inertial range, $\mathcal{F}(k) \sim 1/k$ in YO theory, but $\mathcal{F}(k) = 0$ in Kolmogorov theory.

In an earlier work, \citet{Forster:PRA1977} employed a similar RG analysis for different forcing functions, and derived  various $E(k)$'s, including $E(k) \sim k^2$ for 3D.  We skip these derivations due to lack of space.


\subsection{McComb and Zhou}
\label{sec:RecursiveRG}

McComb, Zhou, and coworkers~\cite{McComb:PRA1983,Zhou:PRA1988} performed self-consistent recursive RG analysis and computed the renormalized viscosity.  They assumed the forcing to be  at large scales, hence noise renormalization is not required in the inertial range, where power-law solution is attempted for the renormalized viscosity.

In the absence of noise, the Feynman diagrams for  McComb and Zhou's RG procedure are same as that employed for  DIA.  McComb and Zhou  performed only coarse-graining, and not rescaling, that leads to an increase in   the renormalized viscosity increases  to
\be
\nu^{(n)}(k)=\nu^{(n+1)}(k)+\delta\nu^{(n)}(k).
\label{eq:nu_n_Zhou}
	\ee
For $\nu^{(n)}(k)$, McComb and Zhou employed power-law solution along with a universal function $\nu_{(n)}^{*}(k')$:
\be
	\nu^{(n)}(k_{n}k')=(K_\mathrm{Ko})^{1/2}\Pi^{1/3}k_{n}^{-4/3}\nu^{(n)*}(k').
	\label{eq:dnu_Zhou}
\ee
Substitution of Eqs.~(\ref{eq:nu_n_Zhou}, \ref{eq:dnu_Zhou}) in Eq.~(\ref{eq:nu_k_second_integral}) yields
\begin{eqnarray}
	\delta\nu^{(n)*}(k') & = & \frac{1}{(d-1)}\int^\Delta d\mathbf{p}'\frac{2}{(d-1)S_{d}}\frac{E(q')}{q'^{d-1}}\left[\frac{Q(k',p',q')}{\nu^{(n)*}(hp')p'^{2}+\nu^{(n)*}(hq')q'^{2}} \right]\label{eq:delta_nu*}, \label{eq:duk_star_Zhou}
	\\
	\nu^{(n+1)*}(k') & = & h^{4/3} \nu^{(n)*}(hk')+h^{-4/3}\delta \nu^{(n)*}(k'),
\end{eqnarray}
where $Q(k',p',q')$ is given in Eq.~(\ref{eq:Q_DIA}), ${\bf{p'+q'=k'}}$, and $h = 1/c$ with $c=k_{n+1}/k_n$ as the coarse-graining parameter.

\begin{figure}
	\begin{center}
		\includegraphics[scale = 0.6]{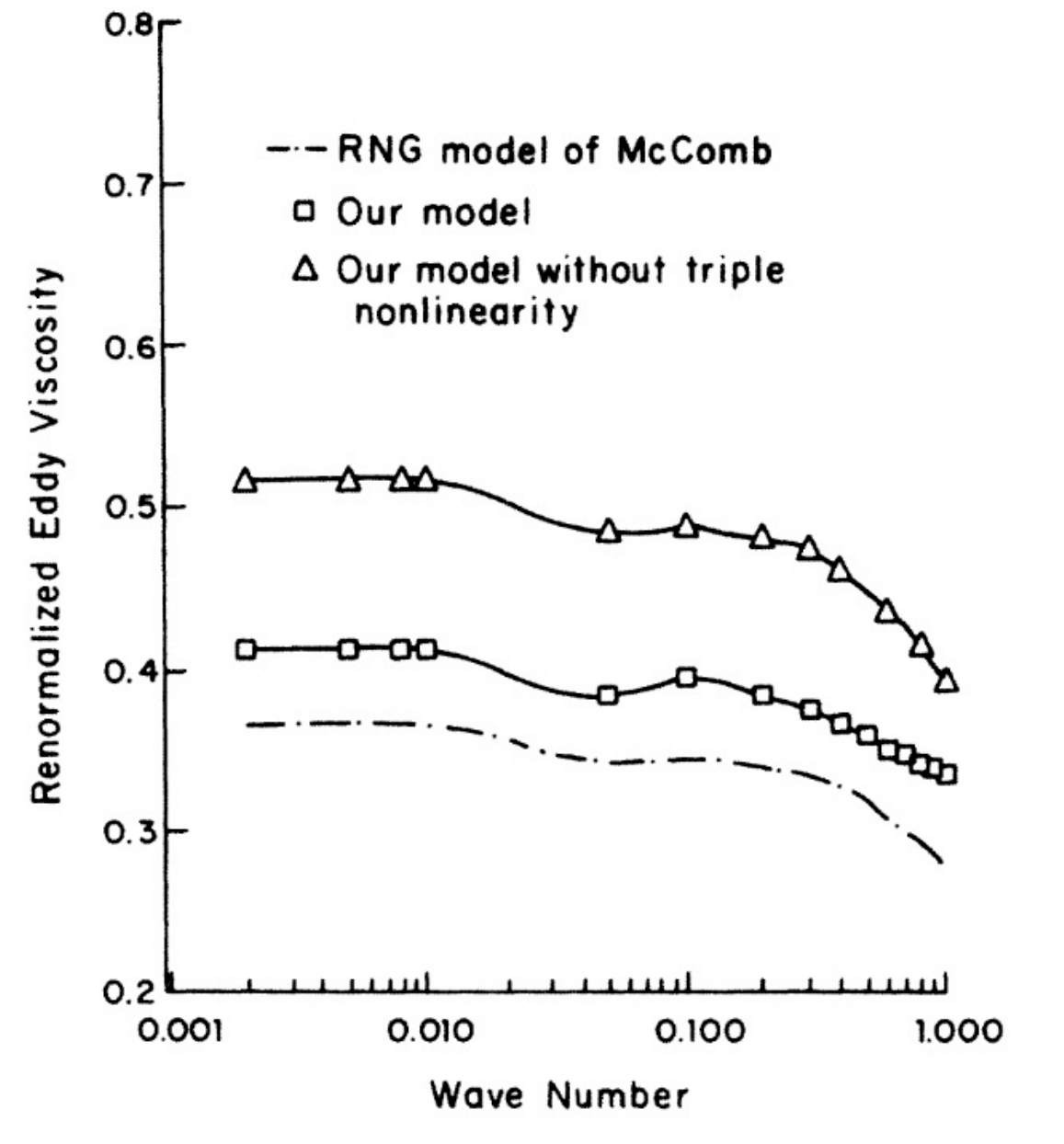}
	\end{center}
	\vspace*{0pt}
	\caption{Plot of $\nu_{(n)}^{*}(k')$ vs.~$k'$ computed  by~\citet{Zhou:PRA1988} without $\la u^< u^< u^< \ra$ (solid line with squares) and with $\la u^< u^< u^< \ra $ (solid line with triangles), and by McComb (chained line).  From  \citet{Zhou:PRA1988}. Reprinted with permission from APS. }
	\label{fig:nu_k_Zhou}
\end{figure}

The $d{\bf p'}$ integral of Eq.~(\ref{eq:duk_star_Zhou}) is computed in the wavenumber band $k_n \le p < k_{n+1}$ and $k_n \le q < k_{n+1}$. McComb, Zhou, and coworkers solved the above equations numerically that yields $\nu_{(n)}^{*}(k')$ shown in Fig.~\ref{fig:nu_k_Zhou}.    \citet{Zhou:PRA1988} reported that $\nu_{(n)}^{*}(k') \rightarrow 0.4$ tor small $k'$, whereas \citet{McComb:PRA1983} reported that $\nu_{(n)}^{*}(k') \rightarrow 0.37$. \citet{Zhou:PRA1988} argued that  an inclusion of triple nonlinearity, $\la u^< u^< u^<\ra$, alters $\nu_{(n)}^{*}(k')$ marginally (see Fig.~\ref{fig:nu_k_Zhou}); but this issue is beyond the scope of this paper. 

\subsection{RG Analysis in Craya-Herring Basis}
\label{sec:RG_CH}
RG analysis is quite complex involving intricate tensor algebra and integrals.  Recently, \citet{Verma:PRE2024_d_dim} employed the Craya-Herring (CH) basis that simplifies the algebra significantly. In addition,    similarities between hydrodynamic, scalar, and magnetohydrodynamic (MHD) turbulence become apparent in CH basis. We will discuss these similarities in Sections~\ref{sec:PS}  and \ref{sec:MHD}.

\citet{Verma:PRE2024_d_dim}  focussed on the  triadic interactions in the CH basis. As shown in Fig.~\ref{fig:CH_triad}, for a triad, the $u_1$ components reside in the plane of the triad, whereas the $u_2$ components are perpendicular to the plane. \citet{Verma:PRE2024_d_dim}  showed that the  renormalized viscosities of the $u_1$ and $u_2$ components are different, which  has many important implications. For example, this feature leads to a conclusion that $d=6$ is the critical upper dimension for  HDT~\cite{Verma:PRE2024_d_dim}. 

\begin{figure}
	\begin{center}
		\includegraphics[scale = 0.7]{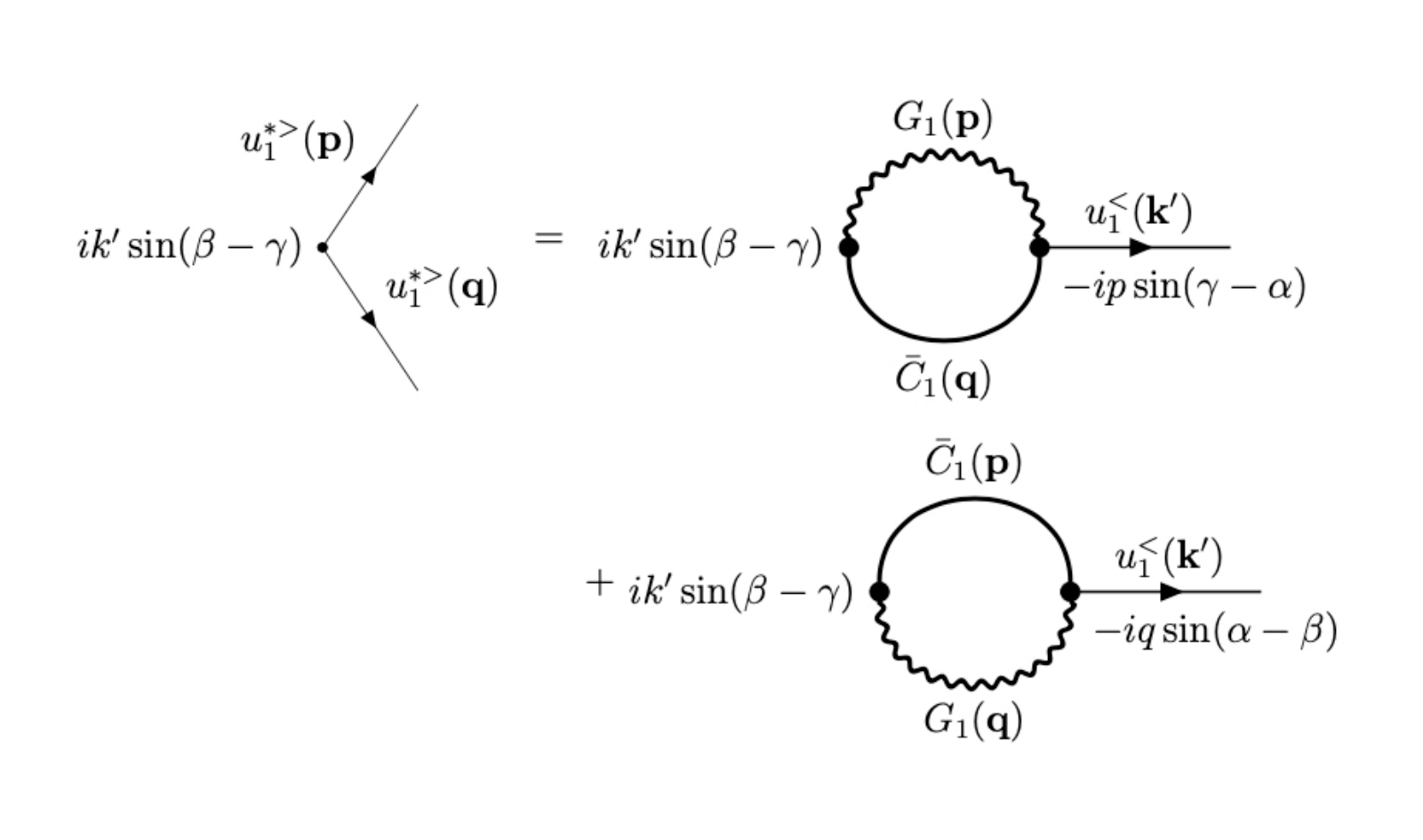}
	\end{center}
	\vspace*{0pt}
	\caption{Feynman diagrams associated with the renormalization of $\nu_1$ for the $u_1$ component. From \citet{Verma:PRE2024_d_dim}.  Reprinted with permission from APS.
	}
	\label{fig:RG_CH1}
\end{figure}

\begin{figure}
	\begin{center}
		\includegraphics[scale = 0.7]{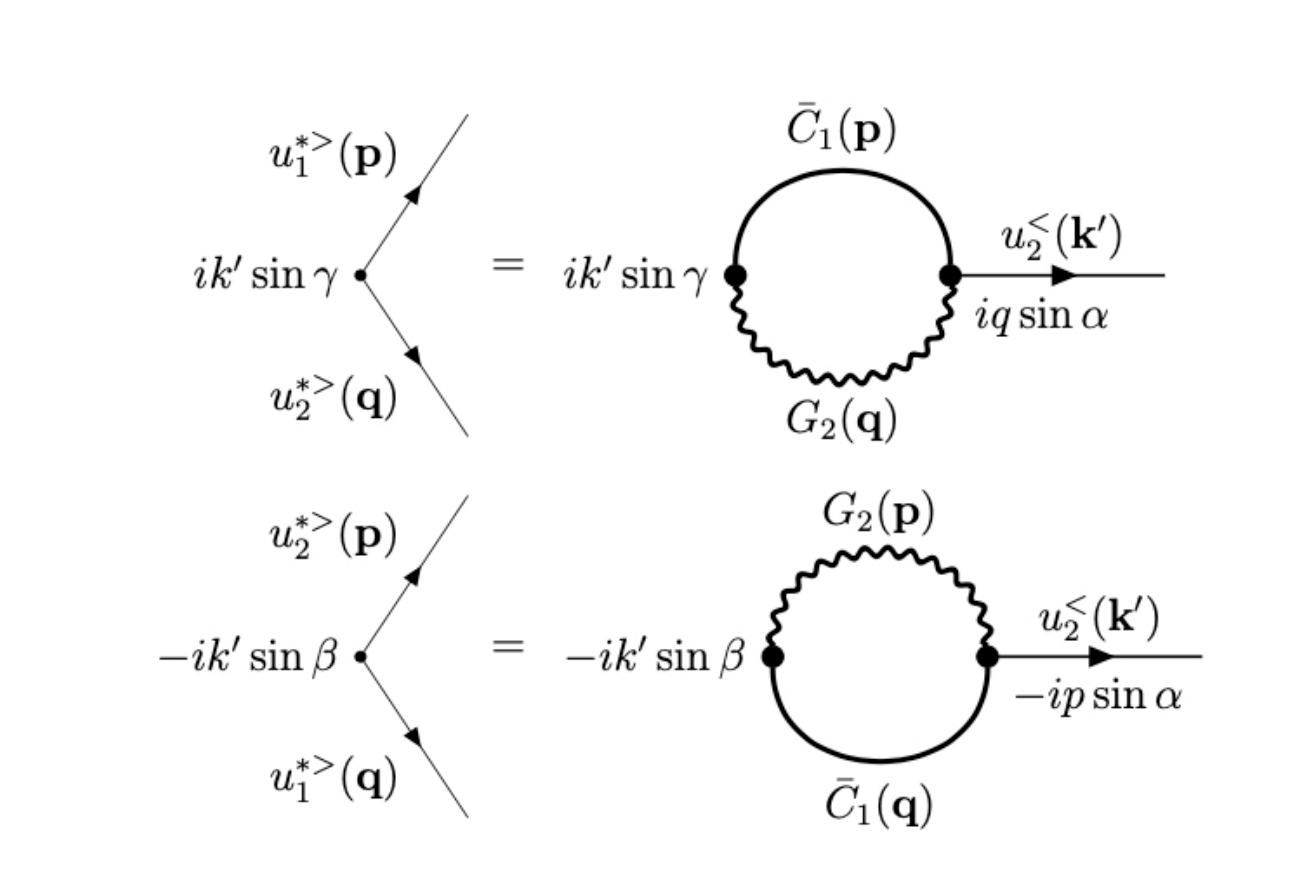}
	\end{center}
	\vspace*{0pt}
	\caption{Feynman diagrams associated with the renormalization of $\nu_2$ for the $u_2$ component.  Modified version of Fig. 7 of \citet{Verma:PRE2024_d_dim}.  Reprinted with permission from APS.}
	\label{fig:RG_CH2}
\end{figure}

In this subsection, we briefly describe the renormalization of the viscosities $\nu_1$ and $\nu_2$ (for $u_1$ and $u_2$, respectively)~\cite{Verma:PRE2024_d_dim}. We start  with Eqs.~(\ref{eq:u1k_dot}, \ref{eq:u2k_dot}) and  compute the viscosity corrections arising due to the nonlinear terms. The Feynman diagrams for the  $\nu_1$ and $\nu_2$ corrections are illustrated in Fig.~\ref{fig:RG_CH1} and Fig.~\ref{fig:RG_CH2}, respectively.  These computations lead to
\bea
\nu^{(n)}_{1} k^2 & = &   \nu^{(n+1)}_1 k^2 - \mathrm{Integrals~corresponding~to~Fig.~\ref{fig:RG_CH1}},
\label{eq:dnu1_k_CH} \\
\nu^{(n)}_{2} k^2 &  =  & \nu^{(n+1)}_2 k^2 - \mathrm{Integrals~corresponding~to~Fig.~\ref{fig:RG_CH2}},
\label{eq:dnu2_k_CH}
\eea
or
\bea
\nu^{(n)}_1 k^2 & = &  \nu^{(n+1)}_1 k^2 -  \int_\Delta \frac{d{\bf p}} {(2\pi)^{d}} \frac{k \sin(\beta-\gamma)  }{\nu_1(p)p^2 + \nu_1(q) q^2} [p C_1({\bf q}) \sin(\gamma-\alpha) +  q C_1({\bf p}) \sin(\alpha-\beta) ], \label{eq:nu1_RG_CH} \nonumber \\ \\
\nu^{(n)}_2 k^2 & = &  \nu^{(n+1)}_2 k^2   -  \int_\Delta \frac{d{\bf p}} {(2\pi)^{d}} 
\left[ \frac{kq C_1({\bf p}) \sin \gamma \sin \alpha 
}{\nu_1(p)p^2 + \nu_2(q) q^2} 
+ \frac{ kp C_1({\bf q}) \sin \beta \sin \alpha 
}{\nu_2(p)p^2 + \nu_1(q) q^2}  \right].
\label{eq:nu2_RG_CH}
\eea 

Following Kolmogorov's theory of turbulence, we model $\nu^{(n)}_1 $ and $\nu^{(n)}_2$ as follows:
\bea
\nu^{(n)}_1 & = & \nu_{1*} \sqrt{K_\mathrm{Ko}} \Pi^{1/3} k_n^{-4/3},  \label{eq:nu1_CH} \\
\nu^{(n)}_2 & = & \nu_{2*} \sqrt{K_\mathrm{Ko}} \Pi^{1/3} k_n^{-4/3}.  \label{eq:nu2_CH} 
\eea
Substitution of Eqs.~(\ref{eq:nu1_CH}, \ref{eq:nu2_CH}) and Kolmogorov's energy spectrum [Eqs.~(\ref{eq:Ek_Ck}, \ref{eq:Ek_Kolm})] in Eqs.~(\ref{eq:nu1_RG_CH}, \ref{eq:nu2_RG_CH}) yields
\bea
\nu_{1*}  (1- b^{-4/3})  &= &-  \frac{2 S_{d-1}}{(d-1) S_d} \frac{1}{\nu_{1*}} \int_1^b p'^{d-1} dp'  \int^{p'/2}_{(p'^2+1-b^2)/(2p')}  dz (1-z^2)^{\frac{d-3}{2}} (F_1+F_2), \nonumber \\
\label{eq:nu1_integral_CH} \\
\nu_{2*}  (1-b^{-4/3}) & = &  \frac{2 S_{d-1}}{(d-1) S_d} \int_1^b p'^{d-1} dp'  \int^{p'/2}_{(p'^2+1-b^2)/(2p')}    dz (1-z^2)^{\frac{d-3}{2}}   F_3(p',z) ,
\label{eq:nu2_integral_CH}
\eea
where  
\bea
F_1(p',z) & = & \frac{(1-z^2)  (p'-2z)(2p'z-1) p' q'^{-8/3-d}}{p'^{2/3} +q'^{2/3}}, \\
F_2(p',z)  & = & \frac{(1-z^2)  (1-p'^2)(2p'z-1) p'^{-2/3-d} q'^{-2}}{p'^{2/3} +q'^{2/3}}, \\
F_3(p',z) & = &  \frac{(1-z^2)  p'^{-2/3-d}}{\nu_{1*} p'^{2/3} + \nu_{2*} q'^{2/3}}
+ \frac{(1-z^2) p'^2 q'^{-8/3-d}}{\nu_{2*} p'^{2/3} + \nu_{1*} q'^{2/3}}.
\eea
See \citet{Verma:PRE2024_d_dim} for details. 

For  $c=1.5$, the integrals yield  $\nu_{1*} = 0.098$ and $\nu_{2*} = 0.62$ for 2D HDT, and  $\nu_{1*} = 0.070$ and 
$\nu_{2*} = 0.53$  for  3D HDT~\cite{Verma:PRE2024_d_dim}. These constants are reasonably close to those reported by \citet{McComb:book:Turbulence} and \citet{Zhou:PRA1988}. \citet{Verma:PRE2024_d_dim} also showed that 
$\nu_{1*} $ vanishes for $d=6$, which leads to the upper  critical dimension for HDT to be 6. That is, the velocity field is Gaussian for $d \ge 6$~\cite{Verma:PRE2024_d_dim}.

\subsection{Functional RG}

In statistical field theory, \textit{partition function} is computed using functional integral of \textit{order parameter}.  Similarly,   quantum field theory   employs functional integrals for many computations. The above  computations are mostly for equilibrium fields. Fortunately, functional integral has been applied to turbulence with some modifications in the standard framework~\cite{DeDominicis:PRA1979,Canet:JFM2022,Antonov:RNC2025}.  

\citet{DeDominicis:PRA1979} started with the following equations:
\bea
\partial_t u_\alpha & = & \nu_0 \nabla^2 u_\alpha + \lambda_0 \tau_{\alpha \beta} ({\bf u} \cdot \nabla) u_\beta + f_\alpha, \\
\tau_{\alpha \beta} & = & \delta_{\alpha \beta} - \frac{k_\alpha k_\beta}{k^2}, 
\eea
for which the \textit{generating functional} was proposed as
\be
\hat{Z}(l) = \int Du D\hat{u} \exp(\mathcal{L}[u, \hat{u}] )
+ \int dt d^4 x l_\alpha(x,t) u_\alpha(x,t)
\ee
with the Lagrangian  as
\be
\mathcal{L} = \int dt d^4  \left[ -i \hat{u}_\alpha (\partial_t - \nu_0 \nabla^2) u_\alpha  - i \hat{u}_\alpha \lambda_0  \tau_{\alpha \beta} ({\bf u} \cdot \nabla) u_\beta + i \hat{u}_\alpha	\la f_\alpha f_\beta \ra \hat{u}_\beta
\right].
	\ee
Here, $\hat{u}$ is the auxiliary function to $u$. Refer to \citet{Antonov:RNC2025} for details.

\citet{DeDominicis:PRA1979} showed that the unperturbed propagators are
\bea
\la i \hat{u}_\alpha u_\beta \ra & = & 
\frac{1}{-i \omega+  \nu_0 k^2} \delta_{\alpha \beta}, \\
\la u_\alpha u_\beta \ra & = & 
\frac{2D_0 k^{4-d}}{|-i \omega+  \nu_0 k^2|^2}  (m_0^2 +k^2)^{-y/2} \tau_{\alpha \beta}(k).
\eea
After some algebra, the relationship between the correlation and Green's functions was derived as
\be
\la u u \ra = C(k, \omega)
= \frac{D_0}{\nu_0^2} G\left(\frac{\omega}{i \nu_0}, k; g_0, \Lambda, m_0 \right)
\ee
with
\be
\frac{\lambda_0^2 D_0}{\nu_0^3} = g_0 \Lambda^y.
\ee
Using Callan-Symanzik equation, \citet{DeDominicis:PRA1979} derived that
\be 
E(k) \sim k^{-y+1+\eta_\nu -\eta_D}.
\ee 
\citet{DeDominicis:PRA1979} showed that for small $y$ and $d>2$, $\eta_D = 0$ and $\eta_\nu  = y/3$.  For $y=4$,  $E(k) \sim k^{-5/3}$.

We end this section with a  summary of RG analysis of HDT.  RG solves the infrared divergence problem encountered in DIA. In addition, RG predicts both the equilibrium solution ($k^{d-1}$ spectrum) and nonequilibrium solution ($k^{-5/3}$ spectrum) for HDT.  Note that RG analysis yields the latter solution even though the coupling constant is $O(1)$. RG computations also show that $d_c =6$ is the critical dimension for HDT; for $d>d_c$, the velocity field is Gaussian, as in $\phi^4$ theory for $d>4$~\cite{Wilson:PR1974,
	Goldenfeld:book}.

In Section~\ref{sec:ET}, I will describe field-theoretic calculations of the energy transfers and flux in HDT.

\section{Energy Transfers and Flux}
\label{sec:ET}

In this section, we  compute the energy fluxes in HDT, both in 2D and 3D.  Here,  nonzero fluxes arise due to the nonequilibrium nature of the flow.

\subsection{Basic Formulas}

The dynamical equation for the \textit{modal energy}   $E({\bf k}) = |{\bf u(k)}|^2/2$ is~\cite{Kraichnan:JFM1959,Leslie:book}
\bea
(\partial_t +2 \nu k^2) E(\mathbf{k},t) & = &  \frac{1}{2} \sum_{\bf p,q}  S^{uu}({\bf k|p,q})  + \Re[ {\bf F}_u({\bf k},t) \cdot {\bf u}^*({\bf k}, t) ]   ,
\label{eq:Euk_combined}
\eea
where
\begin{enumerate}
	\item  $\Re[ {\bf F}_u({\bf k},t) \cdot {\bf u}^*({\bf k}, t) ] $ represents energy injection to the wavenumber \textbf{k}.
	
	\item $2 \nu k^2 E(\mathbf{k},t) $ represents the viscous dissipation rate for the mode $E(\mathbf{k},t)$.
	
	\item  The term   \be
S^{uu}({\bf k|p,q}) = \Im \left[   \{ {\bf k} \cdot  {\bf u}({\bf q}, t) \}   \{ {\bf u}({\bf p}, t) \cdot  {\bf u}^*({\bf k}, t) \} \}  +  \{ {\bf k} \cdot  {\bf u}({\bf p}, t) \}   \{ {\bf u}({\bf q}, t) \cdot  {\bf u}^*({\bf k}, t) \} \}  
\right]
\label{eq:Suu_combined}
\ee
is the \textit{combined energy transfer}   from ${\bf u}({\bf q})$ and ${\bf u}({\bf p})$ to ${\bf u}({\bf k})$~\cite{Kraichnan:JFM1959}. 
\end{enumerate}

Now, let us consider a wavenumber sphere of radius $R$. The external force injects energy into the sphere at the rate of $\sum_{k<R} \Re[ {\bf F}_u({\bf k},t) \cdot {\bf u}^*({\bf k}, t) ] $. In Kolmogorov's theory of turbulence, the energy is injected at large scales \textit{only}. In 3D, nonlinearity helps transfer this energy to    small scales where it is dissipated by the viscous term. An absence of forcing in the inertial range leads to a constant energy flux in the inertial range~\cite{Frisch:book,Verma:JPA2022}~[see Eq.~(\ref{eq:VEF})].  This phenomenology has some similarities with  Yakhot-Orszag~\cite{Yakhot:JSC1986}'s RG predictions when $\epsilon=4$.  For this case, \citet{Yakhot:JSC1986} obtained   nearly constant energy flux and $k^{-5/3}$ energy spectrum in the inertial range (see Section~\ref{sec:YO}).

The energy flux $ \Pi(R)$ for the wavenumber sphere of radius $R$ is the net nonlinear energy transfer rate from all the modes residing inside the sphere to the modes outside the sphere.  Using $S({\bf k|p,q})$, \citet{Kraichnan:JFM1959} derived the following formula for $ \Pi(R)$~\cite{Leslie:book}:
\be
\Pi(R) = \frac{1}{2} \int_R^\infty dk' \int \int^\Delta dp dq S({\bf k'|p,q})-\frac{1}{2} \int_0^R dk' \int \int^\Delta dp dq S({\bf k'|p,q}),
\label{eq:flux_Kr}
\ee
where ${\bf k = p+q}$, and $\Delta$ represents a range of ${\bf p,q}$ consistent with the definition of energy flux~\cite{Kraichnan:JFM1959,Leslie:book}.

\citet{Dar:PD2001} and \citet{Verma:PR2004,Verma:book:ET} showed that 
\be
S^{uu}({\bf k|p|q}) = \Im \left[   \{ {\bf k} \cdot  {\bf u}({\bf q}, t) \}   \{ {\bf u}({\bf p}, t) \cdot  {\bf u}^*({\bf k}, t) \} \}    \right]
\label{eq:Suu}
\ee
is the \textit{mode-to-mode energy transfer rate} from the \textit{giver} mode ${\bf u(p)}$ to the \textit{receiver} mode ${\bf u(k)}$ with the mediation of mode ${\bf u(q)}$.    The combined energy transfer to \textbf{k} from \textbf{p} and \textbf{q} is a sum of $S^{uu}({\bf k|p|q})$ and $S^{uu}({\bf k|q|p})$.   In terms of $S^{uu}({\bf k|p|q})$, the  formula for $  \Pi(R)  $ is
\be
  \Pi(R)   = \int_{R}^\infty \frac{d{\bf k'}}{(2\pi)^d} \int_0^{R} \frac{d{\bf p}}{(2\pi)^d}  S^{u u}({\bf k'|p|q}) .
\label{eq:fluid_flux}
\ee
In Eq.~(\ref{eq:fluid_flux}), the giver modes are within the sphere, whereas the receiver modes are outside the sphere.  In contrast,  \citet{Kraichnan:JFM1959}'s derivation of Eq.~(\ref{eq:flux_Kr}) is  longer and more complex.  Note that Eqs.~(\ref{eq:flux_Kr}, \ref{eq:fluid_flux}) yield the same energy flux for a given velocity field~\cite{Verma:PR2004,Verma:book:ET}.

In  Craya-Herring basis~\citep{Verma:book:ET,Verma:PRE2024_d_dim},
\bea
S^{uu}({\bf k'|p|q}) & = & S^{u_1 u_1}({\bf k'|p|q}) +
S^{u_2 u_2}({\bf k'|p|q}),
\label{eq:Suu_kpq_CH}
\eea
where
\bea
S^{u_1 u_1}({\bf k'|p|q})& = &k' \sin\beta \cos \gamma  \Im \{u_1({\bf q},t) u_1({\bf p},t) u_1({\bf k'},t) \} ,  \label{eq:Su1u1} \\
S^{u_2 u_2}({\bf k'|p|q}) & = &- k' \sin\beta \Im \{u_1({\bf q},t) u_2({\bf p},t) u_2({\bf k'},t) \}. \label{eq:Su2u2}
\eea
Thus, there are independent energy transfers along the $u_1$ and $u_2$ channels, with  no cross transfers between them.

\subsection{Field-theoretic Computation of Energy Flux}
\label{sec:HDT_ET}

Experiments and numerical simulations reveal that $\la 
S^{uu}({\bf k'|p|q}) \ra \ne 0$ for HDT. Nonvanishing  $S^{uu}({\bf k'|p|q})$ (a triple correlation) indicates non-Gaussian and nonequilibrium nature of a turbulent flow.  Interestingly, quasi-normal approximation for the triple correlation yields constant energy flux in the inertial range of HDT~\cite{Kraichnan:JFM1959,Orszag:CP1973,Leslie:book}. I briefly present this derivation in this subsection.

Under the quasi-normal approximation,  the triple correlation of $ S^{uu}({\bf k'|p|q}) $ vanishes to the zeroth order. However, the first-order expansion of the triple correlation yields a sum of fourth-order correlations, which are expanded as respective sums of products of two second-order correlations (assuming Gaussian distribution).  Kolmogorov's  energy spectrum and the renormalized viscosity are employed to compute the flux integrals. 

 \citet{Kraichnan:JFM1959}  (also see \citet{Leslie:book}) was first to compute the turbulence flux using  the above framework.   By expanding Eq.~(\ref{eq:flux_Kr}) to first order, Kraichnan derived the following formula for 3D HDT:
\bea
1 = \frac{ \la \Pi(R)\ra}{\Pi }   = \frac{1}{K_\mathrm{Ko}^{3/2}}  \int_0^1 dv   [\ln(1/v)] \int_{v^*}^{1+v} dv   \Sigma(v,w), 
\label{eq:Pi_Kr_field}
\eea
where $v_* = \mathrm{max}(v,|1-v|)$, and
\be
 \Sigma(v,w) = vw\frac{\{b(1,v,w) w^{-11/3} (v^{-11/3}-1) + b(1,w,v) v^{-11/3} (w^{-11/3}-1) \}} {1+v^{2/3} + w^{2/3}}
  \ee
  with
  \be
  b(1,v,w) = \frac{p}{k} (xy+z^3)
  \ee
  for 3D. The above derivation employs the following transformation~\cite{Kraichnan:JFM1959,Leslie:book}
\be
k = \frac{R}{u};~~p =\frac{Rv}{u};~~q = \frac{Rw}{u}.
\label{eq:uvw_transformation_ET}
\ee
\citet{Kraichnan:JFM1959} and \citet{Leslie:book} computed the  integral of Eq.~(\ref{eq:Pi_Kr_field}) numerically and reported the Kolmogorov's constant for 3D HDT to be near 1.6. Refer to ~\citet{Leslie:book}   for details.   \citet{Verma:PR2004,Verma:arxiv2005} performed similar computations using the mode-to-mode energy transfers~\cite{Verma:PR2004}.

\begin{figure}
	\begin{center}
		\includegraphics[scale = 0.60]{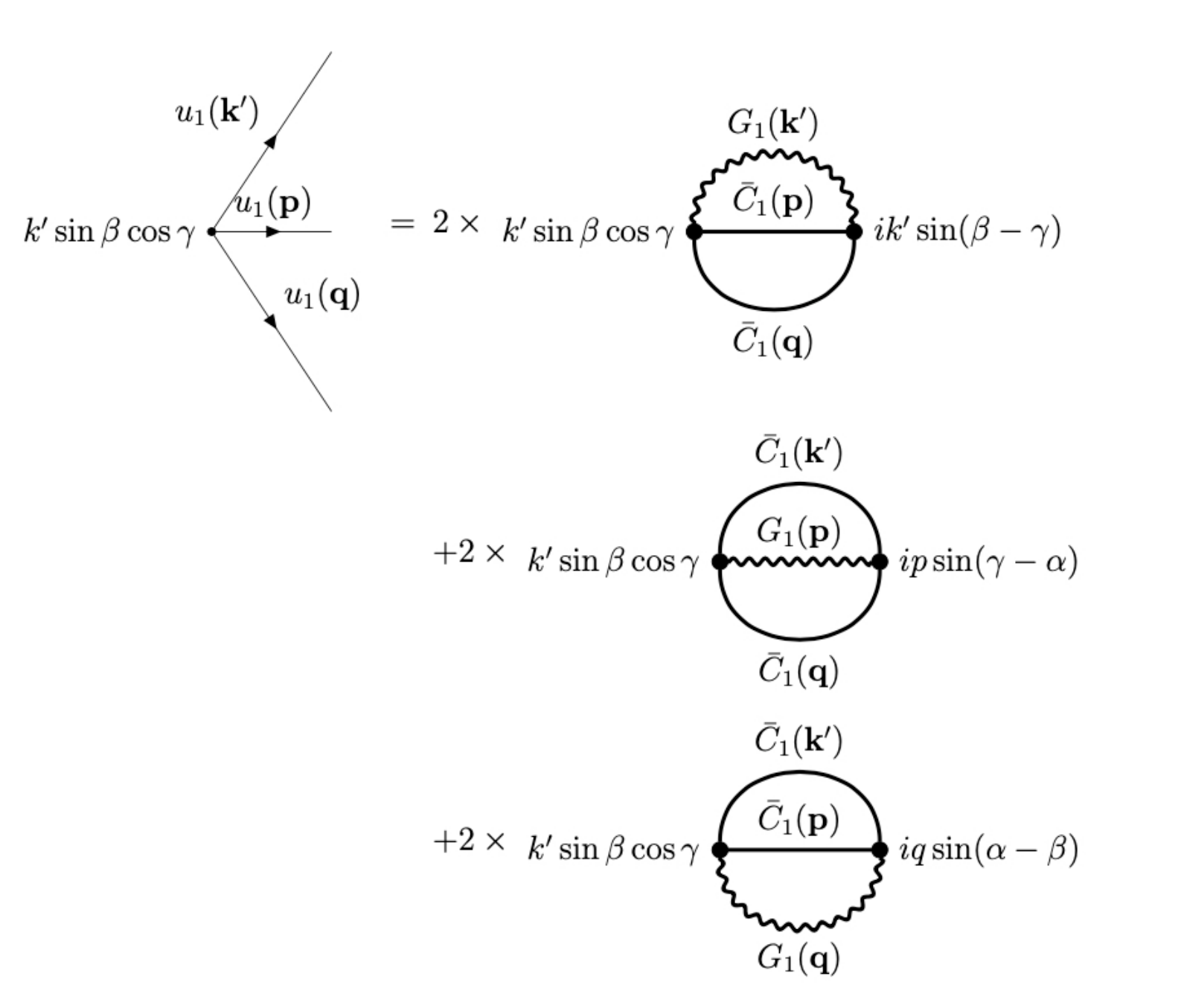}
	\end{center}
	\vspace*{0pt}
	\caption{Feynman diagrams associated with the energy transfers between the $u_1$ components. From \citet{Verma:PRE2024_d_dim}.  Reprinted with permission from APS. }
	\label{fig:ET_CH1}
\end{figure}

In the following discussion, we present the energy flux computations in the Craya-Herring basis~\cite{Verma:PRE2024_d_dim}. Starting from Eq.~(\ref{eq:Su1u1}), \citet{Verma:PRE2024_d_dim} derived the following equation for the $u_1$ component:
\bea
\la S^{u_1 u_1}({\bf k'|p|q}) \ra & = &     \frac{\mathrm{numr}_1 }{\nu_1(k) k^2 + \nu_1(p) p^2 + \nu_1(q) q^2},
\label{eq:Skpq_u1_expanded}
\eea
where
\bea
\mathrm{numr}_1 & = & 2 [k' \sin(\beta-\gamma) C_1({\bf p})C_1({\bf q})+p \sin(\gamma-\alpha) C_1({\bf k'} ) C_1({\bf q})+q \sin(\alpha-\beta) C_1({\bf k'} ) C_1({\bf p}) ]\nonumber \\
&& \times k' \sin\beta \cos \gamma .
\eea
The three terms of $\la S^{u_1 u_1}({\bf k'|p|q}) \ra$ correspond to the three Feynman diagrams  of Fig.~\ref{fig:ET_CH1}.  
\begin{figure}
	\begin{center}
		\includegraphics[scale = 0.8]{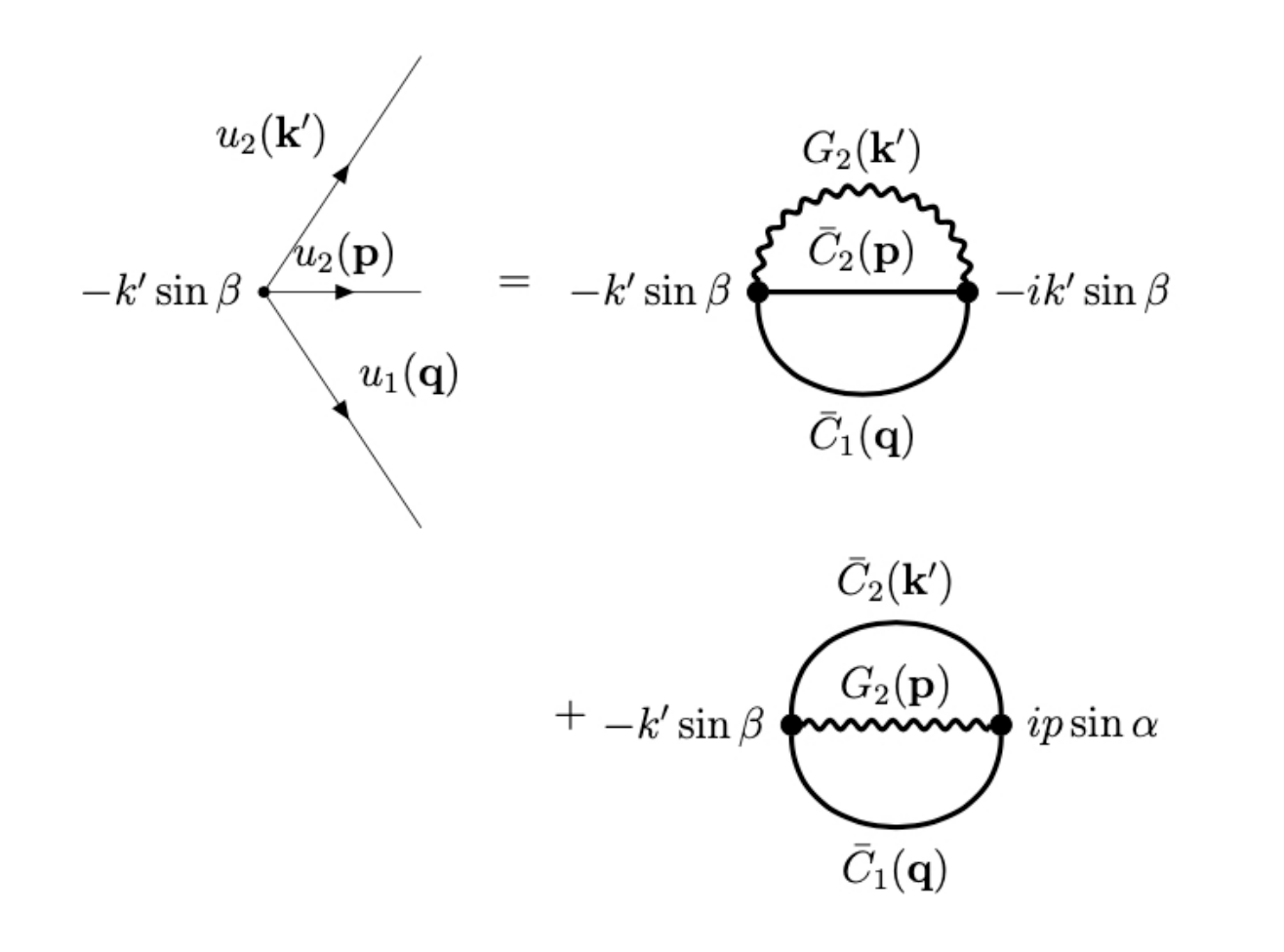}
	\end{center}
	\vspace*{0pt}
	\caption{Feynman diagrams associated with the energy transfers between the $u_2$ components. Modified version of Fig. 9 of \citet{Verma:PRE2024_d_dim}.  Reprinted with permission from APS.}
	\label{fig:ET_CH2}
\end{figure}
For 3D and two-dimension three-component (2D3C), the energy transfers along the $u_2$ channel is
\bea
\la S^{u_2 u_2}({\bf k'|p|q}) \ra  & = & (k \sin\beta)^2   \frac{C_1({\bf q}) [C_2({\bf p}) -C_2({\bf k'})  ] }{\nu_2(k) k^2 + \nu_2(p) p^2 + \nu_1(q) q^2}.
\label{eq:Skpq_u2_expanded}
\eea
The corresponding Feynman diagrams are shown in Fig.~\ref{fig:ET_CH2}. 

Using Eqs.~(\ref{eq:Skpq_u1_expanded}, \ref{eq:Skpq_u2_expanded}), we derive the corresponding energy fluxes:
\be 
\la \Pi_{u_j}(R)\ra = \int_R^\infty \frac{d{\bf k'}}{(2\pi)^d}
\int_0^R \frac{d{\bf p}}{(2\pi)^d} \la S^{u_j u_j}({\bf k'|p|q}) \ra,
\ee
where $j$ takes values 1 or 2. Using the transformations of Eq.~(\ref{eq:uvw_transformation_ET}), \citet{Verma:PRE2024_d_dim}  derived the following nondimensionalized equation for $\la \Pi_{u_j}(R)\ra$:
\bea
\frac{ \la \Pi_{u_j}(R)\ra}{\Pi }   = A  \int_0^1 dv   [\log(1/v)] v^{d-1} \int_{-1}^1 dz (1-z^2)^{\frac{d-3}{2}} \la S^{u_j u_j}(v,z) \ra, 
\label{eq:Pi_integral_CH}
\eea
where
\bea
A & = &  K_\mathrm{Ko}^{3/2}
\frac{4}{(d-1)^2} \frac{S_{d-1}}{S_d} .
\label{eq:A}
\eea
For 2D, the energy flux is $ \Pi_{u_1}(R)$, whereas in 2D and  2D3C,  the total energy flux is 
\be
\Pi(R) =  \Pi_{u_1}(R)+ \Pi_{u_2}(R).
\ee
See \citet{Verma:PRE2024_d_dim}  for details.

\citet{Verma:PRE2024_d_dim} computed the energy fluxes for 2D and 3D turbulence. For 2D, using $\Pi_{u_1}(R) = -\Pi$ and $\nu_{1*} =  0.098$, Verma deduced that $K_\mathrm{Ko} = 1.19$, which is inconsistent with numerical simulations and experiments, according to which $K_\mathrm{Ko} \approx 6$, and with field-theoretic work of  \citet{Nandy:IJMPB1995}, who reported that $K_\mathrm{Ko} = 6.447$.  These differences are possibly due to the inability of the recursive RG schemes to capture the nonlocal interactions.  \citet{Verma:PRE2024_d_dim}  proposed a workaround by changing $\int_0^1 dv $ of Eq.~(\ref{eq:Pi_integral_CH}) to $\int_{0.22}^1 dv $ that yields $K_\mathrm{Ko} = 4.46$.

For 3D HDT, $\nu_{1*}$ and $\nu_{2*}$ appear in the denominator of Eq.~(\ref{eq:Skpq_u1_expanded}, \ref{eq:Skpq_u2_expanded}).  Since $\nu_{1*} \ll \nu_{2*}$, the negative energy flux $\Pi_{u_1}$ dominates positive $\Pi_{u_2}$ leading to  $\Pi(R) <0$, which is inconsistent with the experimental and numerical observations.  Fortunately, this issue is easily resolved by employing $\int_{0.22}^1 dv $ for $\Pi_{u_1}(R)$ of Eq.~(\ref{eq:Pi_integral_CH})  that yields $K_\mathrm{Ko} =1.63$, which is in good agreement with earlier field-theoretic computations, as well as numerical and experimental results~\cite{Verma:PRE2024_d_dim}.  

The  above nonzero energy flux (nonequilibrium) breaks  detailed balance via directional energy transfers---from  small wavenumber modes  to  large wavenumbers.  In contrast, the equilibrium solution, Gaussian velocity field of the spectrally-truncated Euler equation, respects detailed balance and yields vanishing energy flux. Note that the triple correlations of Eqs.~(\ref{eq:Suu}, \ref{eq:Skpq_u1_expanded}, \ref{eq:Skpq_u2_expanded}) vanish following the Gaussian  \textbf{u} of the Euler turbulence~\cite{Verma:book:ET,Verma:PRE2024_d_dim}.  This is consistent with the fact that
 $\la S^{u_j u_j}({\bf k'|p|q}) \ra$'s of  Eqs.~(\ref{eq:Skpq_u1_expanded}, \ref{eq:Skpq_u2_expanded})  vanish for equipartitioned $C_1({\bf k})$ and $C_2({\bf k})$.  Thus, field theory yields the energy transfers for the equilibrium flows (Euler turbulence) and nonequilibrium flows (HDT).

We end this section with a brief-comment on $d$-dimensional turbulence. \citet{Fournier:PRA1978} employed eddy-damped quasi-normal Markovian
(EDQNM) procedure to HDT and observed a transition from a positive  energy flux to a negative energy flux near $d  = 2.05$ as $d$
decreases from 3 to 2. \citet{Adzhemyan:JPA2008}  employed perturbative field theory and showed that the Kolmogorov constant
$K_\mathrm{Ko} \propto d^{1/3}$, which leads to a decrease in the energy flux with the increase in $d$. \citet{Clark:JFM2022} observed   a  transition to a non-chaotic regime above critical dimension $d_c \approx 5.88$.  Recently, \citet{Verma:PRE2024_d_dim} showed that $\nu_{1*} \rightarrow 0$ as $d \rightarrow 6$, and identified $d_c =6$ as the critical dimension of HDT. The works by \citet{Clark:JFM2022} and \citet{Verma:PRE2024_d_dim}  indicate that the velocity field becomes Gaussian for $d \approx 6$ and beyond. Field-theoretic work of \citet{Adzhemyan:JPA2008} is consistent with the above conclusion.

 Many realistic flows are accompanied by scalars (e.g., pollutants, temperature field) or vectors (e.g., magnetic field) or tensors (e.g., polymers).  Field-theoretic treatment of such fields have yielded interesting results~(see e.g., \cite{Adzhemyan:book:RG,Zhou:PR2010,Verma:PR2004}). This topic is extensive and its complete overview is beyond the scope of this review (see \cite{Leslie:book,McComb:book:Turbulence,Sagaut:book,Zhou:PR2010,Zhou:PR2021} for discussion). Yet, for illustration, we briefly describe two cases: a passive scalar advected by the velocity field, and a specific case of magnetohydrodynamic (MHD) turbulence.

\section{Field-theoretic Treatment of Passive Scalar}
\label{sec:PS}
In this section, I discuss a field-theoretic calculation of passive scalar turbulence. In this system, the momentum equation is not coupled to the scalar field. Hence, the equation for the velocity field is same as Eq.~(\ref{eq:NS}). The scalar field $\psi$ is advected by the velocity field, hence its equation is~\cite{Leslie:book,Lesieur:book:Turbulence}
\bea 
\frac{\partial{\psi}}{\partial t} + ({\bf u}\cdot\nabla){\psi}
& = &   \kappa \nabla^2 {\psi} + f_\psi,  
\label{eq:scalar_real}
\eea
where $f_\psi$ is the large-scale force on $\psi$, and $\kappa$ is the scalar diffusion coefficient. 

Equation~(\ref{eq:scalar_real}) is transformed to Fourier space. In the CH basis,  the evolution equation for ${\psi}({\bf k'})$ is~\cite{Verma:book:ET}
 \bea
\left( \frac{\partial}{\partial t} + \kappa k^2 \right) {\psi}({\bf k'}) & = &i k' \int \frac{d {\bf p}}{(2\pi)^d} \{ \sin \gamma u_1^*({\bf p})  \psi^*({\bf q}) -\sin\beta u_1^*({\bf q})  \psi^*({\bf p})\} + f_\psi({\bf k}'), \nonumber \\  \label{eq:scalar:psi_k_dot} 
\eea 
where ${\bf k'+p+q} = 0$. Since Eq.~(\ref{eq:scalar:psi_k_dot}) is same as that  for $u_2({\bf k})$ with $u_2 \rightarrow \psi$, we expect the RG and the energy-transfer equations for the scalar field to be similar to that for $u_2$. This observation provides very useful insights. 

For the scalar field, the correlation function at wavenumber ${\bf k}$ is 
\be 
C^\psi({\bf k}) =  |{\psi({\bf k})}|^2 ,
\ee
and the total scalar energy is
\bea 
\frac{\la \psi^2 \ra}{2} = \int E^\psi(k) dk = \frac{1}{2} \int  \frac{d{\bf k }} {(2\pi)^{d}} C^\psi({\bf k}) = \frac{1}{2} \frac{S_d}{(2\pi)^d} \int dk  k^{d-1} C^\psi({\bf k}), 
\eea
where $E^\psi(k)$ is the one-dimensional (1D) shell spectrum. Using the above equation we derive
\be
E^\psi(k) = \frac{1}{2}  \frac{S_d  k^{d-1} }{ (2\pi)^d} C^\psi({\bf k}).
\label{eq:Ek_scalar}
\ee
The above $E^\psi(k)$ differs from 1D kinetic energy spectrum of Eq.~(\ref{eq:Ek_Ck}) by a factor of $1/(d-1)$, which has  important implications, as we show below. 

For large Reynolds number $UL/\nu \gg 1$, the kinetic energy spectrum $E(k)$ is given by Eq.~(\ref{eq:Ek_Kolm}). For the turbulent velocity field, the passive scalar spectrum depends on the P\'{e}clet number, $UL/\kappa$, that represents the ratio of the nonlinear term ${\bf u} \cdot \nabla \psi$ and the diffusion term~\cite{Gotoh:book_chapter:passive_scalar}. For large Reynolds and P\'{e}clet numbers,  \citet{Obukhov:1949Scalar} and \citet{Corrsin:APL1951} proposed that the passive scalar spectrum is 
\bea 
E^\psi(k) & = & K_\psi \Pi_\psi \Pi^{-1/3} k^{-5/3},  \label{eq:Epsi_CH}
\eea
where $\Pi_\psi $ is the \textit{scalar energy flux}, and $K_\psi $ is \textit{Obukhov-Corrsin constant}. Using atmospheric boundary layer data,  \citet{Champagne:JAS1977}  showed that $K_\psi \approx 0.64$.  \citet{Sreenivasan:PF1995} summarized many past results and argued that 	$K_\psi \approx 0.4$. 

There are a number of papers on field theory of passive scalar turbulence.  Here, we review only a couple of them for $d=3$. \citet{Yakhot:JSC1986} extended their RG computation (Section~\ref{sec:YO}) to passive scalar and reported that   $K_\psi = 1.16$. \citet{Zhou:PRE1993RG_scalar} employed the recursive RG   [see Section~\ref{sec:RecursiveRG}] to the scalar field, and reported that the turbulent Prandtl number ($\nu(k)/\kappa(k)$) is near 0.7. \citet{McComb:JFM1992} employed \textit{local energy transfer} (LET) theory (similar to recursive RG) and obtained $K_\mathrm{Ko} = 2.5$ and $K_\psi = 1.1$. \citet{Antonov:RNC2025} employed functional RG and reported that turbulent Prandtl number is 0.7040.  \citet{Adzhemyan:PRE1998} employed \textit{operator product expansion} to compute anomalous scaling for the passive scalar turbulence (also see \cite{Adzhemyan:book:RG}). \citet{Verma:IJMPB2001} employed  recursive RG  and reported that  the turbulent Prandtl number is 0.42, and  $K_\psi = 1.25$.

\begin{figure}
	\begin{center}
		\includegraphics[scale = 0.5]{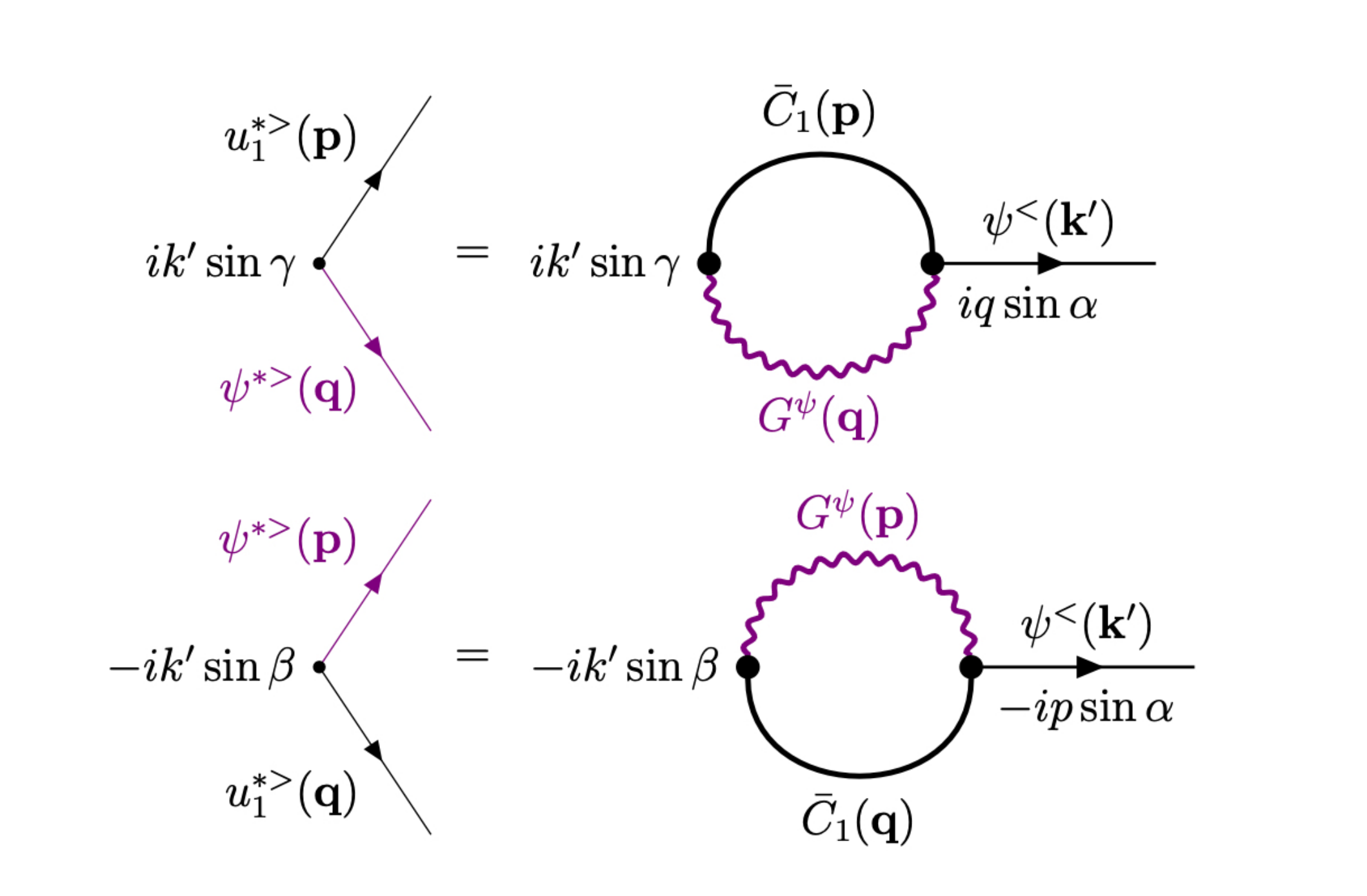}
	\end{center}
	\vspace*{0pt}
	\caption{Passive scalar turbulence: Feynman diagrams associated with  the computation of renormalized diffusivity ($\kappa$).  Here, $G^\psi({\bf k})$ is the Green's function of the scalar field.}
	\label{fig:scalar_RG}
\end{figure}

The CH basis brings out the connections between the HDT and passive scalar turbulence quite nicely, which is  described below.   Since the scalar field does not affect the velocity field, $\nu^{(n)}_1$ follows the same equation as Eq.~(\ref{eq:nu1_CH}).  We derive the renormalized diffusivity following the same procedure as that for $u_2$ (see Section~\ref{sec:RG_CH}).  Figure~\ref{fig:scalar_RG} depicts the Feynman diagrams associated with the first-order perturbation of Eq.~(\ref{eq:scalar:psi_k_dot}). The RG analysis yields
\bea
\kappa^{(n)} k^2 &  =  & \kappa^{(n+1)} k^2 - \mathrm{Integrals~corresponding~to~Fig.~\ref{fig:scalar_RG}},
\label{eq:dkapps_k_CH}
\eea
or
\bea
\kappa^{(n)} k^2 &  =  & \kappa^{(n+1)} k^2    -  \int^\Delta \frac{d{\bf p}} {(2\pi)^{d}} 
\left[ \frac{kq C_1({\bf p}) \sin \gamma \sin \alpha 
}{\nu_1(p)p^2 + \kappa(q) q^2} 
+ \frac{ kp C_1({\bf q}) \sin \beta \sin \alpha 
}{\kappa(p)p^2 + \nu_1(q) q^2}  \right], 
\label{eq:kappa_RG_1}
\nonumber \\
\eea
where $C_1({\bf p})$ is the spectral correlation for the $u_1$ component.  Note that Eq.~(\ref{eq:kappa_RG_1}) has the same form as Eq.~(\ref{eq:nu2_RG_CH}) with $\nu_2(k) \rightarrow \kappa(k)$. 

The time scale for the scalar field ($\tau^\psi_k)$ is proportional to $(k u_k)^{-1}$. Therefore, we model $\kappa^{(n)}$  as~\cite{Lesieur:book:Turbulence,Verma:IJMPB2001}
\bea
\kappa^{(n)} & = & \kappa_{*} \sqrt{K_\mathrm{Ko}} \Pi^{1/3} k_n^{-4/3},  \label{eq:kappa_CH} 
\eea
which has the same form as Eq.~(\ref{eq:nu2_CH}), but with $\nu_{2*} \rightarrow \kappa_*$. Substitution of the above functions in the RG equation~(\ref{eq:kappa_RG_1})  yields
\bea
\kappa_{*}  (1-b^{-4/3}) & = &  \frac{2 S_{d-1}}{(d-1) S_d} \int_1^b p'^{d-1} dp'  \int^{p'/2}_{(p'^2+1-b^2)/(2p')}    dz (1-z^2)^{\frac{d-3}{2}}   F_4(p',z) , \nonumber \\
\label{eq:kappa_integral} 
\eea
with
\bea
F_4(p',z) & = &  \frac{(1-z^2)  p'^{-2/3-d}}{\nu_{1*} p'^{2/3} + \kappa_{*} q'^{2/3}}
+ \frac{(1-z^2) p'^2 q'^{-8/3-d}}{\kappa_{*} p'^{2/3} + \nu_{1*} q'^{2/3}}.
\label{eq:scalar_integral}
\eea
A comparison of Eq.~(\ref{eq:nu2_integral_CH}) and Eq.~(\ref{eq:scalar_integral}) yields
\be
\kappa_{*} = \nu_{2*}.
\ee
Hence, the renormalized $\kappa$ and renormalized $\nu_2$ are equal, both for 2D and 3D.  Consequently, the turbulent Prandtl number 
\be
\frac{\nu_2(k)}{\kappa(k)}  = \frac{\nu_{2*}}{\kappa_*} =  1.
\ee
  Interestingly, the temporal evolution of  passive scalar in 2D and $u_z$ (velocity component perpendicular to the plane) in 2D3C  are identical [see Eqs.~(\ref{eq:u2k_dot}, \ref{eq:scalar:psi_k_dot})].  Hence, their field theories are identical too.

\begin{figure}
	\begin{center}
		\includegraphics[scale = 0.8]{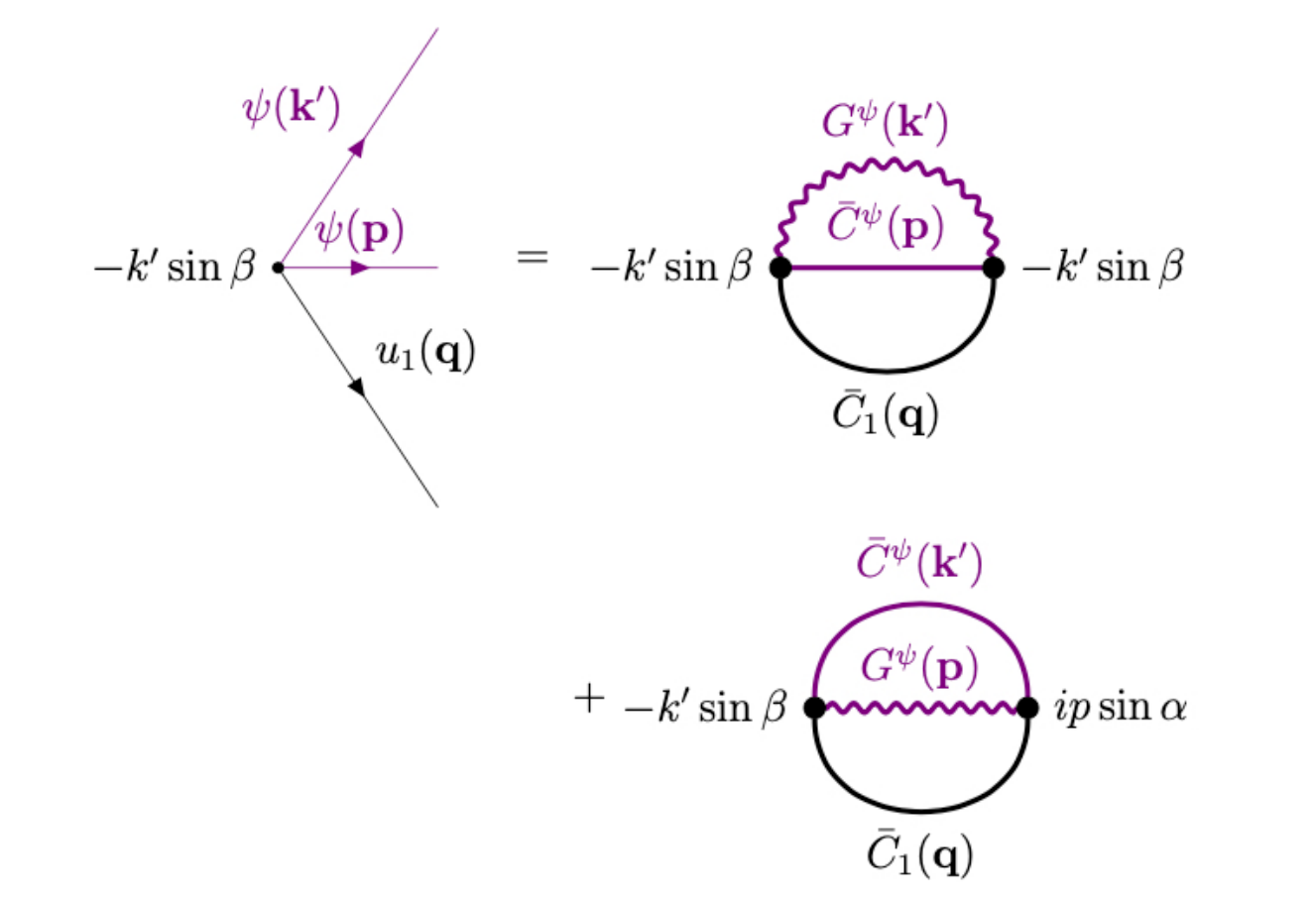}
	\end{center}
	\vspace*{0pt}
	\caption{Passive scalar turbulence: Feynman diagrams associated with  the computation of  mode-to-mode energy transfer  ($\la S^{\psi \psi}({\bf k'|p|q}) \ra $).}
	\label{fig:scalar_ET}
\end{figure}

Now, we compute the scalar energy flux using field theory. The mode-to-mode scalar energy transfer from $\psi({\bf p})$ to  $\psi({\bf k'})$  with the mediation of  ${\bf u}({\bf q})$ is~\cite{Verma:IJMPB2001}
\bea
\la S^{\psi \psi}({\bf k'|p|q}) \ra  = -\Im \left[ \la {\bf  \{  k' \cdot u(q) \} \{ {\psi}({\bf p})  {\psi}({\bf k'}) \} } \ra \right],
\eea
where ${\bf k'+p+q} =0$.  We compute the above triple correlation under quasi-Gaussian assumption. The Feynman diagram for the first-order perturbation, shown in Fig.~\ref{fig:scalar_ET},  is  similar to that  for $u_2$ component in HDT (see Fig.~\ref{fig:ET_CH2}). Following the  procedure outlined in Section~\ref{sec:ET}, we  derive the following expression for $\la S^{\psi \psi}({\bf k'|p|q}) \ra$:
\bea
\la S^{\psi \psi}({\bf k'|p|q}) \ra  & = & (k \sin\beta)^2   \frac{C_1({\bf q}) [C^\psi({\bf p}) -C^\psi({\bf k'})  ] }{\kappa(k) k^2 + \kappa(p) p^2 + \nu_1(q) q^2},
\eea
or
\bea
\la S^{\psi \psi}(v,z) \ra & = &     \frac{v^2 w^{-8/3-d} (v^{-2/3-d}-1) (1-z^2)}{ \kappa_{2*}(1+v^{2/3}) +\nu_{1*}  w^{2/3}},
\label{eq:S_psi(v,w)}
\eea
where $v,w$ are defined in Eq.~(\ref{eq:uvw_transformation_ET}).

Using the above mode-to-mode formula, we derive the corresponding scalar energy flux for a wavenumber sphere of radius $R$ as
\be 
\la \Pi_{\psi}(R)\ra = \int_R^\infty \frac{d{\bf k'}}{(2\pi)^d}
\int_0^R \frac{d{\bf p}}{(2\pi)^d} \la S^{\psi \psi}({\bf k'|p|q}) \ra,
\ee
whose nondimensionalized version is
\bea
\frac{ \la \Pi_{\psi}(R)\ra}{\Pi_\psi }   = A_\psi  \int_0^1 dv   [\log(1/v)] v^{d-1} \int_{-1}^1 dz (1-z^2)^{\frac{d-3}{2}} \la S^{\psi \psi}(v,z) \ra, 
\label{eq:Pi_psi_integral_CH}
\eea
where
\be
A_\psi =  K_\psi K_\mathrm{Ko}^{1/2}
\frac{4}{(d-1)} \frac{S_{d-1}}{S_d} .
\label{eq:Apsi}
\ee
 For $j=2$, Eq.~(\ref{eq:Pi_integral_CH}) has a similar form as Eq.~(\ref{eq:Pi_psi_integral_CH}) with $A \rightarrow A_\psi $ [see Eq.~(\ref{eq:A})]. Also note that $\la \Pi_{\psi}(R)\ra = \Pi_\psi$. Based on these observations, we derive that 
\be
K_\psi = \frac{K_\mathrm{Ko}}{d-1}
\label{eq:Kpsi}
\ee

The factor $d-1$ in Eq.~(\ref{eq:Kpsi}) arises because $E^\psi(k)$ in Eq.~(\ref{eq:Ek_scalar}) lacks the factor $d-1$ in comparison to  Eq.~(\ref{eq:Ek_Ck}). Equation~(\ref{eq:Kpsi}) reveals that $K_\psi  =  K_\mathrm{Ko}/2 \approx 0.8$ in 3D, which is consistent with \citet{Champagne:JAS1977}'s conclusion that  $K_\psi   \approx 0.64$, but inconsistent with \citet{Sreenivasan:PF1996}'s results.  \citet{Verma:IJMPB2001} did not account for the $1/(d-1)$ factor correctly, which led to incorrect $K_\psi$.  Interestingly, in 2D, the integral of Eq.~(\ref{eq:Pi_psi_integral_CH}) is positive, thus indicating a positive  scalar energy flux  despite inverse cascade of kinetic energy (in the $k^{-5/3}$ regime).  Also,  $K_\psi  =  K_\mathrm{Ko}$ in 2D. Theses predictions need to be tested in future numerical simulations. 

The above calculations  demonstrate  usefulness of CH basis in field theory. It provides a unified and transparent framework for HDT and passive scalar turbulence.  In the next section, we employ field theory to MHD turbulence.

\section{Field-theoretic Treatment of MHD Turbulence}
\label{sec:MHD}

Magnetohydrodynamics (MHD) is a description of quasi-neutral magnetized plasma at the continuum level~\cite{Bellan:book:Plasma}. In  Fourier space, the MHD equations are~\cite{Bellan:book:Plasma}
  \bea
\frac{d}{d t}  {\bf u} (\mathbf{k}) + i \int \frac{d {\bf p}}{(2\pi)^d} {\bf \{ k \cdot u(q) \} u(p)} 
& = & -i {\bf k} p (\mathbf{k}) +i \int \frac{d {\bf p}}{(2\pi)^d}   {\bf  \{ k \cdot b(q) \} b(p)}  \nonumber \\
&& -  \nu k^{2}  {\bf u}(\mathbf{k}) + {\bf f}(\mathbf{k}) ,  \label{eq:MHD_formalism:MHDuk}\\
\frac{d}{d t}  {\bf b} (\mathbf{k}) + i  \int \frac{d {\bf p}}{(2\pi)^d}   {\bf \{ k \cdot u(q) \} b(p)}
& = &  i  \int \frac{d {\bf p}}{(2\pi)^d}  {\bf  \{ k \cdot b(q) \} u(p)}  -  \eta k^{2}  {\bf b}(\mathbf{k}) , \label{eq:MHD_formalism:MHDBk}\\
{\bf k\cdot u}(\mathbf{k}) & = & 0 \label{eq:MHD_formalism:k_uk_zero}, \\
{\bf k\cdot b}(\mathbf{k}) & = & 0 \label{eq:MHD_formalism:k_Bk_zero}, 
\eea 
where \textbf{u, b,} and \textit{p} are the velocity, magnetic, and pressure fields, respectively;  $\nu$ and $\eta$ are the kinematic viscosity and magnetic diffusivity, respectively; ${\bf k = p+q}$;  and the external force ${\bf f}$ is employed to the velocity field to maintain  a steady state. In this review, we assume the mean  magnetic field to be absent, which makes the flow statistically isotropic.  We perform field-theoretic analysis using the Els\"{a}sser variables, 
\be 
{\bf z}^\pm = {\bf u \pm b},
\ee
in terms of which the abve MHD equations are
\bea
\frac{d}{d t}  {\bf z}^\pm (\mathbf{k}) + i \int \frac{d {\bf p}}{(2\pi)^d}   {\bf \{ k \cdot z^\mp(q) \} z^\pm(p)} & = & -i {\bf k} p (\mathbf{k})  + \nu_+ k^2 {\bf z}^\pm (\mathbf{k})  + \nu_- k^2 {\bf z}^\mp (\mathbf{k}) + {\bf f}(\mathbf{k}), 
\label{eq:zpm} \nonumber \\ \\
{\bf k\cdot z^\pm}(\mathbf{k}) & = & 0,
\eea
where  $\nu_\pm = (\nu \pm \eta)/2$.  

MHD turbulence has four nonlinear terms involving \textbf{u} and \textbf{b} fields that makes MHD turbulence more complex than HDT. There are several models of MHD turbulence, including anisotropic ones~\cite{Biskamp:book:MHD,Verma:PR2004,Verma:book:ET}.  In this review, we focus on \textit{isotropic MHD turbulence}, in particular on two competing turbulence phenomenologies  whose spectral indices are 3/2 and 5/3 respectively.   \citet{Kraichnan:PF1965MHD}, \citet{Iroshnikov:SA1964}, and \citet{Dobrowolny:PRL1980} argued that the energy spectra for the velocity and magnetic fields are equipartitioned with
\be 
E(k) = E^b(k) =  K_\mathrm{Kr} (\Pi_\mathrm{tot} B_0)^{1/2} k^{-3/2},
\label{eq:Kr_spec}
\ee
where $E(k), E^b(k)$ are the kinetic and magnetic energy spectra, respectively; $\Pi_\mathrm{tot}$ is the total energy flux (a sum of kinetic and energy fluxes);  $K_\mathrm{Kr}$ is Kraichnan's constant; and $B_0$ is the mean magnetic field or an average amplitude of the large-scale magnetic fields.   

Regarding the alternate phenomenology, \citet{Marsch:RMA1991} was the first to propose Kolmogorov-like turbulence phenomenology, according to which the energy spectra $E^\pm(k) $ for $ {\bf z}^\pm$ are (also see \cite{Verma:JGR1996DNS,Verma:PR2004})
\be
E^\pm(k) = K^\pm \frac{(\Pi^\pm)^{4/3}}{(\Pi^\mp)^{2/3}} k^{-5/3},
\ee
where $\Pi^\pm$ are the energy fluxes for  $ {\bf z}^\pm$, and $K^\pm$ are the respective constants.  The spectral exponents 5/3 and 3/2 are quite close, and they are not easily differentiable in numerical simulations or in solar wind observations~\cite{Goldstein:ARAA1995,Verma:PR2004,Schekochihin:JPP2022}. Various authors~\cite{Verma:JGR1996DNS,Mason:PRL2006,Biskamp:PP2000,Beresnyak:PRL2011}  report one exponent or the other, thus making this issue inconclusive from the numerical perspective.  In this section, we do not consider field theory of Kraichnan-Iroshnikov theory\footnote{Kraichnan-Iroshnikov theory is discussed briefly in Section~\ref{sec:WT} in the framework of weak turbulence.}. Instead, we focus on field theory of Kolmogorov-like turbulence phenomenology.

Leading field-theoretic works on MHD turbulence are as follows. \citet{Fournier:JPA1982} generalized  \citet{Forster:PRA1977}'s framework  to MHD turbulence, and predicted various scaling regimes as a function of space dimension $d$ and the ratio of kinetic energy and magnetic energy.  \citet{Adzhemyan:TMP1986} performed similar calculations using quantum field theory renormalization group~\cite{Adzhemyan:book:RG}. \citet{Goldreich:ApJ1995} analyzed anisotropic MHD turbulence under strong turbulence limit, and showed that $E(k_\perp) \sim k_\perp^{-5/3}$ and  $k_\perp z_{k_\perp} \sim k_\parallel B_0$, where $k_\perp, k_\parallel$ are, respectively, the wavenumber components perpendicular and parallel to the mean magnetic field  ${\bf B}_0$. The latter relation is called \textit{critical balance}~\cite{Goldreich:ApJ1995}. Verma~\cite{Verma:PRE2001,Verma:PP1999,Verma:Pramana2003Nonhelical,Verma:Pramana2003Helical,Verma:PR2004} performed RG analysis of $\nu$, $\eta$, and $B_0$ (mean magnetic field renormalization), as well as  energy flux calculations for MHD turbulence; these calculations favor Kolmogorov-like turbulence phenomenology over Kraichnan-Iroshnikov phenomenology.  \citet{Mizerski:JFM2021} employed  RG
 to extract turbulent electromagnetic force for dynamo.  A reader may also refer to reviews~\cite{Verma:PR2004,Schekochihin:JPP2022}.

In the following discussion, we illustrate RG and energy flux computations for a simplified version of MHD turbulence. Here, we work with ${\bf z}^\pm$ variables and  assume that 
\begin{gather} 
\la E^+(k) \ra  =  \la E^-(k) \ra,  \label{eq:Ep_eq_Em}  \\
\Re \la  \left[ {\bf z}^+({\bf k}) \cdot  {\bf z}^{-*}({\bf k})  \right] \ra =  E^u({\bf k})   - E^b({\bf k}) = 0 \label{eq:Er_0}.
\end{gather} 
These assumptions simplify the algebra considerably. For this special case, the total energy (a sum of kinetic and magnetic energies) spectrum $E_\mathrm{tot}(k)  =  E^+(k) = E^-(k)$, and
\be 
E_\mathrm{tot}(k) = K \Pi_\mathrm{tot}^{2/3} k^{-5/3},
\ee
where $K = K^+=K^-$ and $\Pi_\mathrm{tot} = \Pi^+ = \Pi^-$~\cite{Goldstein:ARAA1995,Verma:PR2004}.  Using $\tau_k^\pm \sim (k z^{\mp}_k)^{-1}$, we derive the respective renormalized diffusion coefficients as
\be
\eta_\pm(k) = \eta_{\pm*} \sqrt{K^\mp} \frac{(\Pi^\mp)^{2/3}}{(\Pi^\pm)^{1/3}} k^{-4/3}.
\ee
For the special case discussed above, $\Pi^+ = \Pi^-$ and $\eta_+(k) = \eta_-(k)$. 

\begin{figure}
	\begin{center}
		\includegraphics[scale = 0.7]{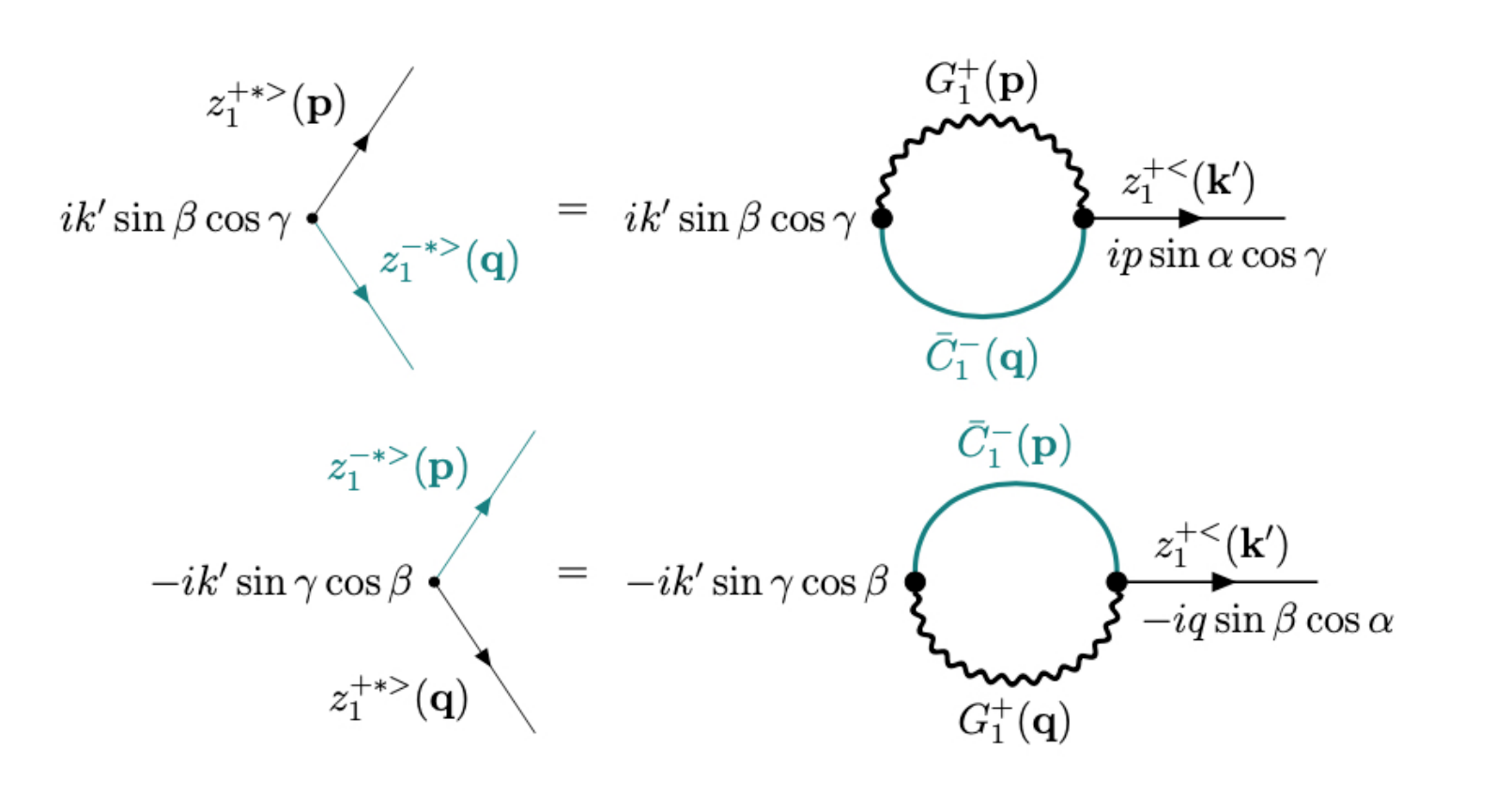}
	\end{center}
	\vspace*{0pt}
	\caption{MHD turbulence: Feynman diagrams associated with  the computation of renormalized diffusivity $\eta_1$. }
	\label{fig:MHDRG_z1}
\end{figure}

\begin{figure}
	\begin{center}
		\includegraphics[scale = 0.5]{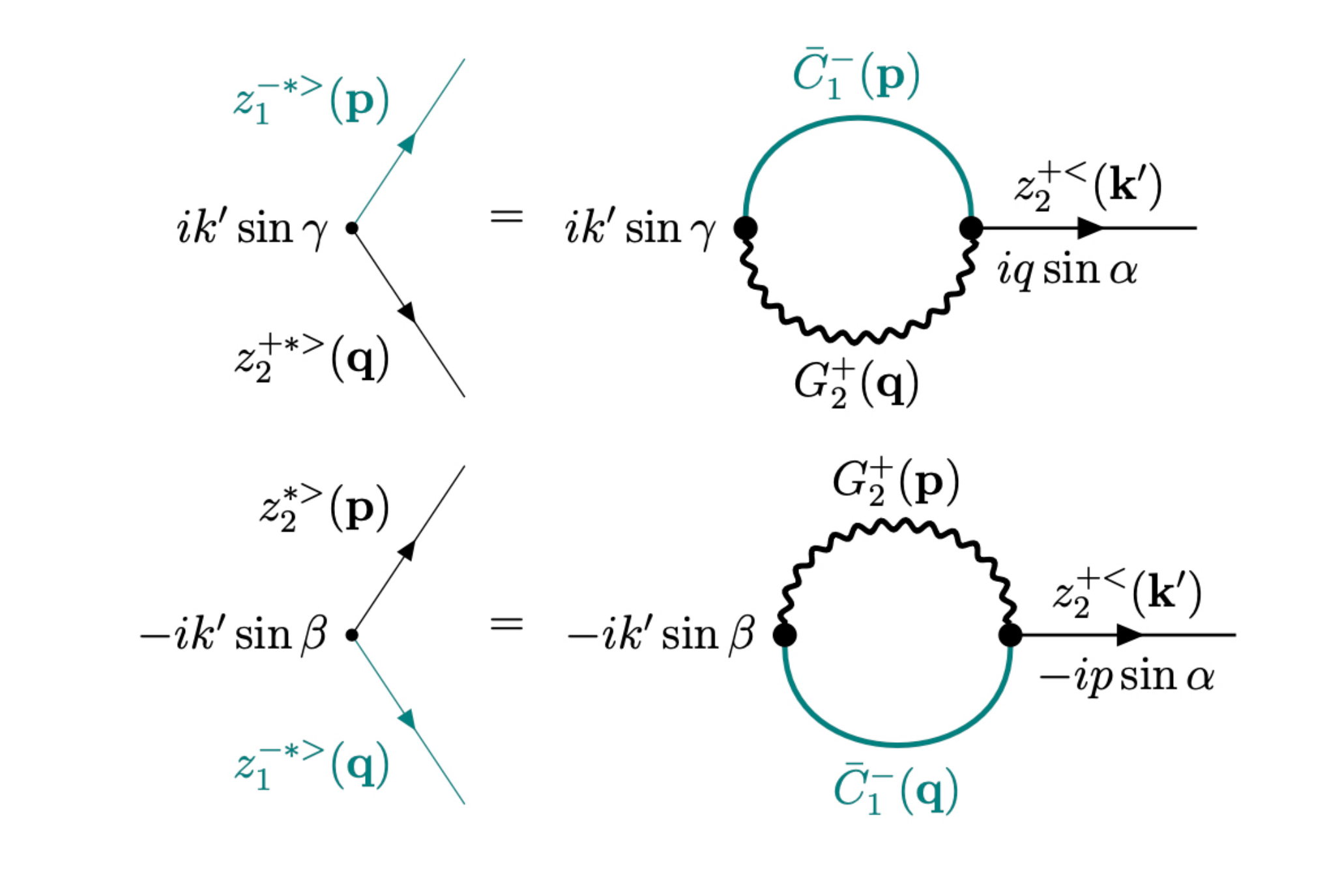}
	\end{center}
	\vspace*{0pt}
	\caption{MHD turbulence: Feynman diagrams associated with  the computation of renormalized diffusivity $\eta_2$. }
	\label{fig:MHDRG_z2}
\end{figure}

As in HDT and passive scalar turbulence, CH basis simplifies the field-theoretic computations of MHD turbulence significantly. In the CH basis, the equations for $z^\pm_1({\bf k'}) $ and $z^\pm_2({\bf k'}) $ are given below:
\bea
\left( \frac{\partial}{\partial t} + \eta_1 k^2 \right)z^\pm_1({\bf k'}) & = & i k' \int \frac{d {\bf p}}{(2\pi)^d} 
[\sin \beta \cos \gamma  z_1^{\mp*}({\bf q}) z_1^{\pm*}({\bf p})   - \sin \gamma \cos \beta z_1^{\mp*}({\bf p}) z_1^{\pm*}({\bf q})  ]  \nonumber \\
&& + f_1(\mathbf{k}'),
\label{eq:vector:w1_dot} \\
\left( \frac{\partial}{\partial t} + \eta_2 k^2 \right)z^\pm_2({\bf k'}) & = &i k'  \int \frac{d {\bf p}}{(2\pi)^d}   \{ \sin \gamma z_1^{\mp*}({\bf p})  z_2^{\pm*}({\bf q}) -\sin\beta z_1^{\mp*}({\bf q})  z_2^{\pm*}({\bf p})\} + f_2(\mathbf{k}').
\label{eq:vector:w2_dot} \nonumber \\
\eea
As in HDT, we assume different diffusive coefficients for $z_1$ and $z_2$ components as follows:
\bea
\eta_{1+} = \eta_{1-} = \eta_1, \\
\eta_{2+} = \eta_{2-} = \eta_2. 
\eea
 To compute the renormalized $\eta_{1,2}$, we expand the nonlinear terms of Eqs.~(\ref{eq:vector:w1_dot}, \ref{eq:vector:w2_dot}) to first order (as in Section~\ref{sec:RG_CH}). The corresponding Feynman diagrams are shown in Figs.~\ref{fig:MHDRG_z1} and  \ref{fig:MHDRG_z2} that yield  the following recurrence relations for the renormalized $\eta_1$ and $\eta_2$:
\bea
\eta^{(n)}_1 k^2 & = &  \eta^{(n+1)}_1 k^2 +  \int_\Delta \frac{d{\bf p}} {(2\pi)^{d}} \frac{1 }{\nu_1(p)p^2 + \nu_1(q) q^2} [k p \sin \beta \sin \alpha \cos^2 \gamma C^-_1({\bf q})  \nonumber \\
& & \hspace{1 in} + k q \sin \beta \sin \gamma \cos^2 \beta C^-_1({\bf p})  ],  \\
\eta^{(n)}_2 k^2 & = &  \eta^{(n+1)}_2 k^2   + \int_\Delta \frac{d{\bf p}} {(2\pi)^{d}} 
\left[ \frac{kq C^-_1({\bf p}) \sin \gamma \sin \alpha 
}{\eta_1(p)p^2 + \eta_2(q) q^2} 
+ \frac{ kp C^-_1({\bf q}) \sin \beta \sin \alpha 
}{\eta_2(p)p^2 + \eta_1(q) q^2}  \right].
\eea 
Note that $C_1^+ = C_1^- = C_1$ because of Eq.~(\ref{eq:Ep_eq_Em}). Also, the vanishing of the cross terms in Eq.~(\ref{eq:Er_0}) help remove some other terms. Thus, Eqs.~(\ref{eq:Ep_eq_Em}, \ref{eq:Er_0}) simplify the RG equations considerably. 


Following similar steps as  for HDT (Section~\ref{sec:RG_CH}), we derive that
\bea
\eta_{(1,2)*}  (1-b^{-4/3}) & = &  \frac{2 S_{d-1}}{(d-1) S_d} \int_1^b p'^{d-1} dp'  \int^{p'/2}_{(p'^2+1-b^2)/(2p')}    dz (1-z^2)^{\frac{d-3}{2}}   F_{(5,6)}(p',z) ,
\nonumber \\
\label{eq:eta_integral}
\eea
where
\bea
F_5(p',z) & = &  \frac{(1-z^2) z^2 p'^2 q'^{-8/3-d}}{\eta_{1*} p'^{2/3} + \eta_{1*} q'^{2/3}}
+ \frac{(1-z^2) (1-p'z)^2 p'^{-8/3-d}/w^2}{\eta_{1*} p'^{2/3} + \eta_{1*} q'^{2/3}} , \\
F_6(p',z) & = & \frac{(1-z^2)  p'^{-2/3-d}}{\eta_{1*} p'^{2/3} + \eta_{2*} q'^{2/3}}
+ \frac{(1-z^2) p'^2 q'^{-8/3-d}}{\eta_{2*} p'^{2/3} + \eta_{1*} q'^{2/3}}.
\eea
The integrals of Eq.~(\ref{eq:eta_integral}) converge for both 2D and 3D.  For the RG parameter $c=1.5$,  $\eta_{1*} = 0.22$ and $\eta_{2*} = 0.50$ for 2D, and  $\eta_{1*} = 0.19$ and  $\eta_{2*} = 0.48$ for 3D. 

\begin{figure}
	\begin{center}
		\includegraphics[scale = 0.5]{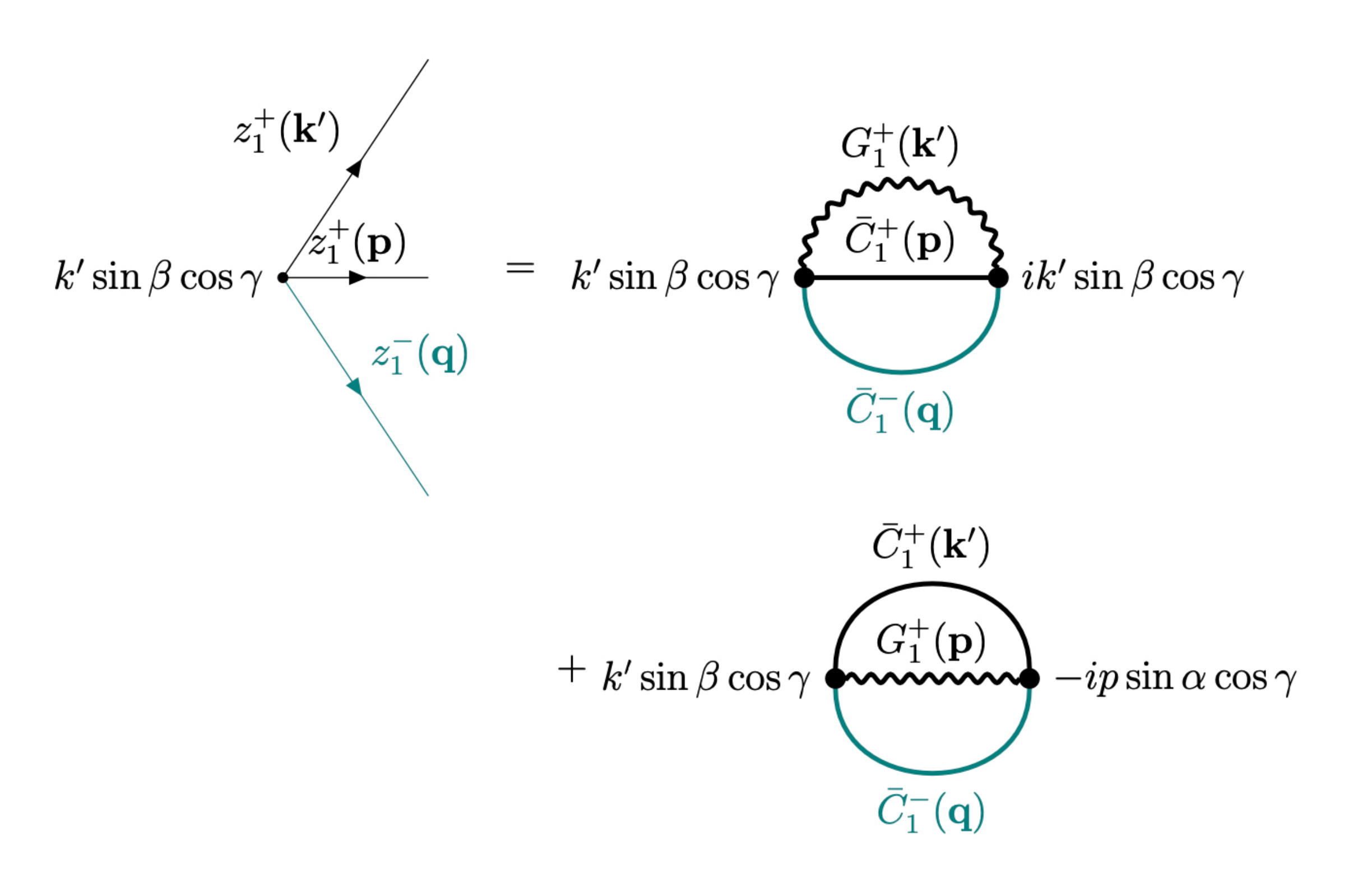}
	\end{center}
	\vspace*{0pt}
	\caption{MHD turbulence: Feynman diagrams associated with  the computation of  mode-to-mode energy transfer via $z_1 z_1$ channel ($\la S^{z_1 z_1}({\bf k'|p|q}) \ra $).}
	\label{fig:MHDET_z1}
\end{figure}

\begin{figure}
	\begin{center}
		\includegraphics[scale = 0.8]{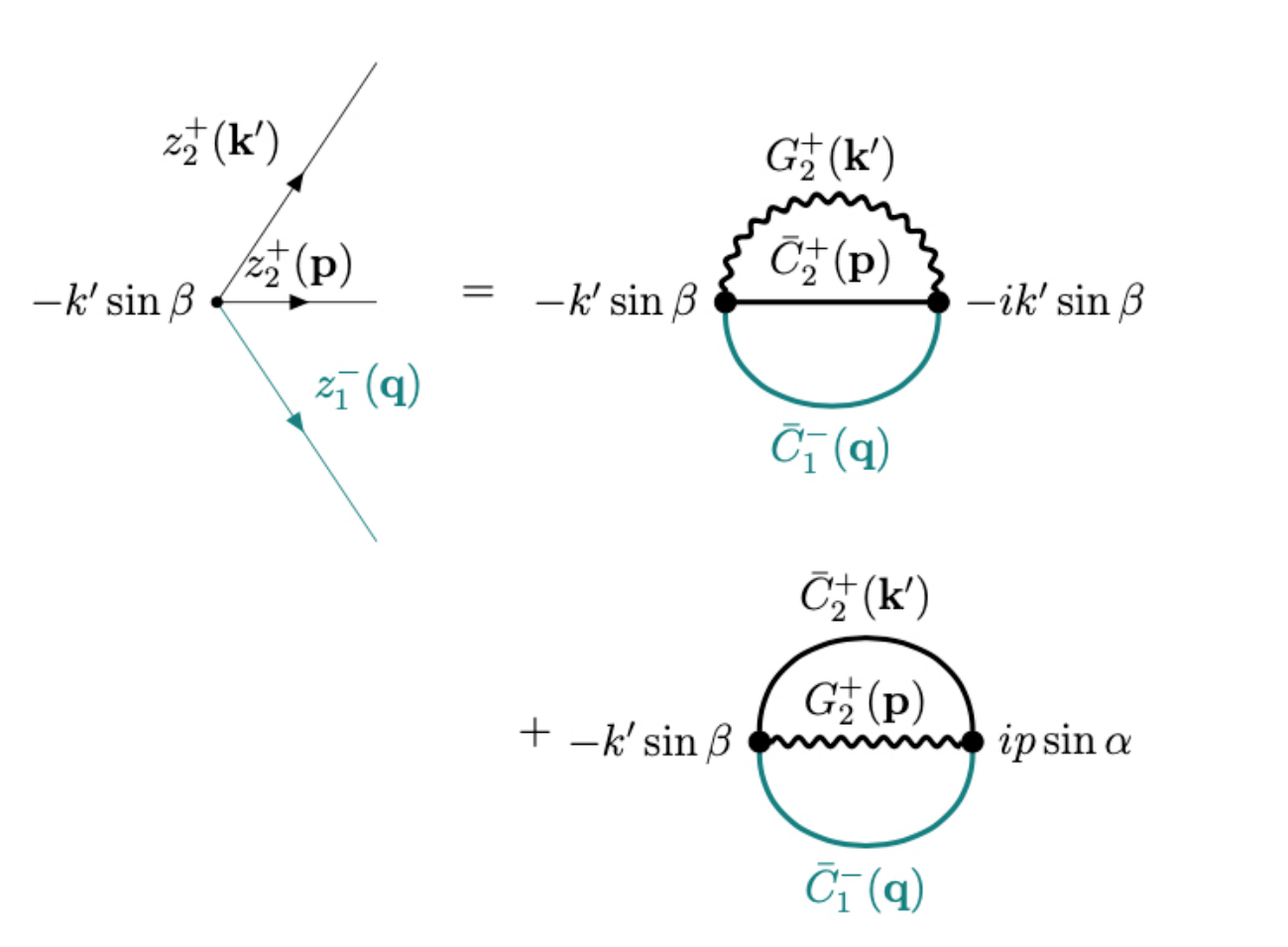}
	\end{center}
	\vspace*{0pt}
	\caption{MHD turbulence: Feynman diagrams associated with  the computation of  mode-to-mode energy transfer via $z_2 z_2$ channel ($\la S^{z_2 z_2}({\bf k'|p|q}) \ra $).}
	\label{fig:MHDET_z2}
\end{figure}

After this, we compute the energy fluxes for ${\bf z}^\pm$ using the following formulae:
\bea
S^{z_1 z_1}({\bf k'|p|q}) & = &  k' \sin\beta \cos \gamma  \Im \{z^-_1({\bf q}) z^+_1({\bf p}) z^+_1({\bf k'}) \} , \label{eq:vector:CH_S11kpq}  \\ 
S^{z_2 z_2}({\bf k'|p|q}) &= & -   k' \sin\beta \Im \{z^-_1({\bf q}) z^+_2({\bf p}) z^+_2({\bf k'}) \}  .  \label{eq:vector:CH_S22kpq}    
\eea 
Since $\la E^+(k) \ra = \la E^-(k) \ra $,  ${\bf z^+}$ and ${\bf z^-}$ channels have the same energy transfers statistically. The above energy transfers vanish to the zeroth order. Hence, we expand  $S^{z_1 z_1}({\bf k'|p|q})$ and $S^{z_2 z_2}({\bf k'|p|q})$ to  first order that yields Feynman diagrams shown in Figs.~\ref{fig:MHDET_z1} and \ref{fig:MHDET_z2} and the following formulas:
\bea
\la S^{z_1 z_1}({\bf k'|p|q}) \ra  & = & (k \sin\beta)^2   (\cos \gamma)^2 \frac{C^-_1({\bf q}) [C^-_1({\bf p}) -C^-_1({\bf k'})  ] }{\eta_1(k) k^2 + \eta_1(p) p^2 + \eta_1(q) q^2}, \\
\la S^{z_2 z_2}({\bf k'|p|q}) \ra  & = & (k \sin\beta)^2   \frac{C^-_1({\bf q}) [C^-_2({\bf p}) -C^-_2({\bf k'})  ] }{\eta_2(k) k^2 + \eta_2(p) p^2 + \eta_1(q) q^2}.
\eea

For 2D, the energy flux $\Pi(R)$  is
\be 
\la \Pi(R)\ra = \la \Pi^+(R)\ra = \la \Pi^-(R)\ra = \int_R^\infty \frac{d{\bf k'}}{(2\pi)^d}
\int_0^R \frac{d{\bf p}}{(2\pi)^d} \la S^{z_1 z_1}({\bf k'|p|q}) \ra.
\ee
Following similar steps as in Sec.~\ref{sec:HDT_ET}, we derive the  constant $K^+=K^- = K = 0.85$ for 2D MHD turbulence. The above relation also indicates that the energy cascades for ${\bf z}^\pm$ are positive in 2D, consistent with the absolute equilibrium theory for MHD turbulence~\cite{Kraichnan:ROPP1980,Stribling:PP1991}. This is in contrast to 2D HDT that exhibits inverse energy cascade.   For 3D MHD turbulence, 
\be 
\la \Pi(R)\ra =  \la \Pi^+(R)\ra = \la \Pi^-(R)\ra = \int_R^\infty \frac{d{\bf k'}}{(2\pi)^d}
\int_0^R \frac{d{\bf p}}{(2\pi)^d} \left[ \la S^{z_1 z_1}({\bf k'|p|q}) \ra + \la S^{z_2 z_2}({\bf k'|p|q}) \ra \right].
\ee
After the integral computation, we obtain $K^+=K^-  = K = 0.96$ for 3D.  

In MHD turbulence, we can vary various parameters: $E^+(k)/E^-(k)$, $E(k)/E_b(k)$, and $\Re[{\bf u(k) \cdot b^*(k)}]$, whose variations lead to significant changes in the turbulence properties, including $K^\pm$ and $\Pi^\pm$.  This is possibly the reason why our field-theoretic results on $K^\pm$ differ from the past works. For example, in 3D, for $\la E^+(k) \ra = \la E^-(k) \ra $  and $\la E(k)/E_b(k) \ra \approx 0.7$, \citet{Verma:PR2004}'s field theory work yields  $K^+ \approx K^- \approx 1.5$. Using numerical simulations,  \citet{Beresnyak:PRL2011}  reported that the Kolmogorov's constant lies between 3.2 to 4.2 depending on Alfv\'{e}nicity.  Note that $\la E^+(k) \ra = \la E^-(k) \ra$ and/or  $\la E(k) \ra = \la E_b(k) \ra$ differ for these works.  We need to carefully compare the  field-theoretic and numerical results for various cases.  However, we do not delve into these details in this review.

In  Section~\ref{sec:WT} we discuss field theory of weak turbulence. 

\section{Field Theory of Weak Turbulence}
\label{sec:WT}

In HDT, scalar turbulence, and MHD turbulence, we compute the renormalized diffusive parameters and  energy fluxes using field theory.  In such flows,  the nonlinearity dominates the linear term, which is not the case for some other systems, for example, weakly nonlinear water waves, strongly rotating flow, MHD with strong ${\bf B}_0$, etc.  For the latter kind of systems, we do not renormalize the parameter(s), but compute the energy  fluxes.  The energy spectrum is deduced using the energy flux formula.  Such a framework is called \textit{weak turbulence}~\cite{Zakharov:book:WaveTurb,Nazarenko:book:WT,Rosenhaus:Arxiv2024}.  In this section, we will briefly illustrate this framework and apply  it to the \textit{advection equation}.

\subsection{Weak Turbulence Framework}

\citet{Zakharov:book:WaveTurb}  extended the  quantum field theory formalism to weak turbulence. We illustrate this framework starting with Schr\"{o}dinger equation:
\be 
i \partial_t \psi = (H_0 +H_1) \psi,
\label{eq:Schrodinger}
\ee
where $\psi$ is the wavefunction, and $H_0$ and $H_1$ are the bare (unperturbed) and perturbed  parts of the Hamiltonian, respectively.  Here, we set  the Planck constant $\hbar=1$. Zakharov assumed $H_1$ to be a nonlinear function of $\psi$, and that $||H_1 || \ll ||H_0||$.  As an example, we consider
\be 
H_0 \psi({\bf k}) = \hbar \omega({\bf k}) \psi({\bf k});
~~H_1 \psi = -\nabla^2 \psi^3.
\label{eq:WT_H0_H1}
\ee 
In Fourier space, Eqs.~(\ref{eq:Schrodinger}, \ref{eq:WT_H0_H1}) transform to the following:
\be 
i \partial_t \psi({\bf k},t) = \omega({\bf k}) \psi({\bf k}, t) 
+k^2  \int \frac{d\bf p}{(2\pi)^d} \frac{d\bf q}{(2\pi)^d} 
\psi({\bf p}, t)  \psi({\bf q}, t)  \psi({\bf s}, t) ,
\label{eq:Schrodinger_k}
\ee 
where ${\bf s =k -p-q}$. Here, the nonlinearity is cubic; it  comes in different order in other systems.

Using Eq.~(\ref{eq:Schrodinger_k}) we  derive the following equation for the particle density $\la |\psi({\bf k},t)|^2 \ra$:
\be 
\partial_t \frac{1}{2} \la  |\psi({\bf k}, t)|^2 \ra = k^2 \Im \left[ \int \frac{d\bf p}{(2\pi)^d} \frac{d\bf q}{(2\pi)^d} \la
\psi({\bf p}, t)  \psi({\bf q}, t) \psi({\bf s}, t)  \psi^*({\bf k}, t) \ra \right] = T_N({\bf k}, t), 
\label{eq:schrodinger_particle}
\ee
where $T_N({\bf k}, t)$ is the nonlinear particle transfer term\footnote{A typical weak turbulence calculation employs $ \psi({\bf k}, t) =  a({\bf k}, t) \exp(-i \omega({\bf k}) t)$, as in the interaction picture of quantum mechanics. We avoid this extra step in our derivation. }. Note that the linear term vanishes in Eq.~(\ref{eq:schrodinger_particle}). Following the discussion of Sec.~\ref{sec:HDT_ET}, we deduce the following formula for the particle flux $\Pi_N(R)$ for a wavenumber sphere of radius $R$:
\be 
\Pi_N(R) = -\int \frac{d\bf k'}{(2\pi)^d} T_N({\bf k'});
\ee
$\Pi_N(R)$ is interpreted as the rate of particle transfer for a wavenumber of radius $R$.  Similar formula can be constructed for the energy flux. 

In a conservative or isolated system [e.g., in Eq.~(\ref{eq:Schrodinger})], the total number of particles [$\int d{\bf k} |\psi({\bf k}, t)|^2$] is conserved, and a wavenumber sphere of radius $R$  contains a finite number of particles. Therefore,  a \textit{steady} particle flux from this sphere is impossible because this flux will deplete the particles in the sphere in a finite time.   Finite $\Pi_N(R)$ is possible only when the particles are constantly injected into the system, which is a nonequilibrium and open framework.  Weak turbulence framework assumes energy and particle injections by an external source (e.g., electromagnetic or mechanical forcing for Bose-Einstein condensate~\cite{Barenghi:PNAS2014}) and dissipation at small scales. We will illustrate a detailed calculation of weak turbulence in the next subsection.   It is important to emphasize  that  a typical isolated or  conservative system \textit{thermalizes  or reaches equilibrium} asymptotically; that is, the energy and particles are evenly distributed among all the Fourier modes.   A quantum system may approach Fermi-Dirac or Bose-Einstein statistics asymptotically depending on the particle nature.

Next, we apply wave turbulence theory to the advection equation.

\subsection{Application of Weak Turbulence to Advection Equation}
\label{subsec:Advection}

The advection equation for a scalar field $\phi$ is
\be 
 \partial _t \phi + V \partial_x \phi = - \partial_x \frac{\phi^2}{2},
 \label{eq:advection}
\ee
where $V$ is the advection speed. Here, we assume that the advection term, $ V \partial_x \phi$ dominates the nonlinear term, $- \partial_x \phi^2/2$. The equation for the modal scalar energy is
\be 
 \partial_t \frac{1}{2} \la |\phi({\bf k},t)|^2 \ra = T({\bf k},t) 
 = \frac{k_x}{2} \int \frac{d\bf p}{(2\pi)^d} \Im  \la  \phi({\bf p},t) \phi({\bf q},t) \phi^*({\bf k},t) \ra ,
 \label{eq:adv_Tk}
\ee 
where ${\bf k = p+q}$, and $T({\bf k},t) $ is the transfer term for the scalar energy. Note  the advection term is cancelled out in Eq.~(\ref{eq:adv_Tk}). In addition, the Green's function and the unequal-time correlation function for the unperturbed advection equation are
\bea
G({\bf k},t-t') & = & H(t-t') \exp[-i k_x V(t-t')], \\
\bar{C}({\bf k},t-t') & = & C({\bf k}) \exp[-i k_x V(t-t')],
\eea
where $C({\bf k})$ is the equal-time correlation function (assumed to be a steady function), and $H(t-t')$ is the heaviside function.  Note that  $V$ is not renormalized because the nonlinear term is much weaker than advection term.

In this review, we compute $T({\bf k},t) $ using the procedure of Sec.~\ref{sec:HDT_ET}.   An expansion of Eq.~(\ref{eq:adv_Tk}) to first-order yields Feynman diagrams similar to Fig.~\ref{fig:ET_CH1}. The energy transfer term corresponding to the first Feynman diagram of Fig.~\ref{fig:ET_CH1} is 
\bea 
T_1 {\bf k},t) & = & \Im \left\{ \frac{i k_x^2}{4} \int \frac{d\bf p}{(2\pi)^d} \int_0^t dt' G({\bf k}, t-t') \bar{C}({\bf p},t-t')  \bar{C}({\bf q},t-t')  \right\} \nonumber \\
& = & \Im \left\{ \frac{i k_x^2}{4} \int \frac{d\bf p}{(2\pi)^d}  
\frac{C({\bf p}) C({\bf q})}{[-i V (k_x -p_x-q_x) + \epsilon]}
 \right\} \nonumber \\
 & = &\frac{\pi k_x^2}{4V}    \int \frac{d\bf p}{(2\pi)^d}  
 \delta(k_x -p_x-q_x)  C({\bf p}) C({\bf q}) ,
 \label{eq:T1_advection}
\eea
where $\epsilon$ is a small parameter that induces dissipation at small scales~\cite{Zakharov:book:WaveTurb}. In the last step, we employed \textit{Cauchy principal value theorem}:
\be 
\frac{1}{x+i\epsilon} = P\left(\frac{1}{x} \right) - i \pi \delta(x),
\ee 
where $P(1/x)$ is the Cauchy principal value.  Following similar steps for the other two Feynman diagrams, we obtain
\be 
T ({\bf k},t) =\frac{\pi k_x}{4V}    \int \frac{d\bf p}{(2\pi)^d}  
\delta(k_x -p_x-q_x)  \left[ k_x C({\bf p}) C({\bf q}) 
- p_x C({\bf q}) C({\bf k}) - q_x C({\bf p}) C({\bf k}) \right].
\ee 

The next task is to compute the inertial-range energy flux, which is
\bea
\Pi(k) = - \int_0^k  \frac{d\bf k'}{(2\pi)^d}   T({\bf k'},t) .
\eea 
The above integral is quite complex with significant anisotropy. Zakharov transformation~\cite{Zakharov:book:WaveTurb} is often used to simplify the flux integration.  I avoid the computation details in this short review. Instead, I simplify the calculation by assuming isotropy,   $p_x \rightarrow p$,  and   $C({\bf p}) \sim E(p)/p^{d-1}$. In addition, as is customary in weak turbulence, I assume \textit{locality}, according to which  wavenumbers with near magnitudes interact ($k \approx p \approx q$), leading to $C(k') \approx C(p) \approx C(q)$~\cite{Zakharov:book:WaveTurb}.  Therefore,
\bea 
\Pi(k) & = & \frac{k^2}{V} k^d k^{d-1} \left( \frac{E(k)}{k^{d-1}} \right)^2 .
	\eea
Hence, for a constant $\Pi(k) = \Pi$,
\be
E(k) \sim (\Pi V)^{1/2} k^{-3/2}.
\label{eq:Ek_3by2}
\ee
Thus, the scalar energy spectrum for the advection equation is $k^{-3/2}$ (not  $k^{-5/3}$). 

Interestingly, for MHD turbulence with strong ${\bf B}_0$, \citet{Kraichnan:PF1965MHD} and \citet{Iroshnikov:SA1964} argued that the advection of  Alfv\'{e}n waves  by ${\bf B}_0$ via  ${\bf B}_0 \cdot \nabla {\bf z}^\pm$  dominates the nonlinear terms, where ${\bf z}^\pm$ are the amplitudes of the Alfv\'{e}n waves in terms of Els\"{a}sser variables.  For ${\bf B}_0 = B_0 \hat{x}$,  
\be 
{\bf B}_0 \cdot \nabla {\bf z}^\pm = B_0 \partial_x {\bf z}^\pm,
\ee 
which has a similar form as the advection term in Eq.~(\ref{eq:advection}).  Multiple fields (${\bf z}^\pm$) and different resonant conditions make the weak MHD turbulence  more complex than advection equation. Still, following the derivation outlined in this subsection, we can derive that $E(k) \sim k^{-3/2}$~\cite{Iroshnikov:SA1964,Nazarenko:book:WT}. A more sophisticated derivation by \citet{Galtier:JPP2000} yields $E(k_\perp) \sim k_\perp^{-2}$; this derivation, however, is beyond the scope of this review.

\subsection{Sweeping Effect and $k^{-3/2}$ Spectrum}

Section~\ref{subsec:Advection} is a gist of Kraichnan's arguments for the $k^{-3/2}$ spectrum for HDT~\cite{Kraichnan:PF1964Eulerian}. Note, however, that Kraichnan~\cite{Kraichnan:PF1964Eulerian} assumed the advection speed $V$ to be random, which leads to 
\bea
G({\bf k},t-t') & = & H(t-t') \la \exp[-i {\bf k \cdot V} (t-t')] \ra =H(t-t') \exp[- \frac{1}{2} V^2 k^2 (t-t') ], \nonumber \\ \\
\bar{C}({\bf k},t-t') & = & C({\bf k}) \la \exp[-i {\bf k \cdot V} (t-t')] \ra =  C({\bf k}) \exp[- \frac{1}{2} V^2 k^2 (t-t') ], 
\eea
substitution of which in Eq.~(\ref{eq:T1_advection})  yields Eq.~(\ref{eq:Ek_3by2}), albeit with a slightly more complex $dt'$ integral in Eq.~(\ref{eq:T1_advection}).  Here, an advection of  fluctuations by $V$ (\textit{random} mean flow or large-scale eddies) yields  $k^{-3/2}$ spectrum in Eulerian framework, which is inconsistent with the experimental observations of $k^{-5/3}$ spectrum. Based on this result, \citet{Kraichnan:PF1964Eulerian} argued that Eulerian approach is inappropriate for the field-theoretic treatment of HDT.  Later, Kraichnan went on to create Lagrangian-based field theory, e.g., \textit{Mixed Lagrangian-Eulerian approach}~\cite{Kraichnan:PF1964Lagrangian_Eulerian}, \textit{Lagrangian-History Closure Approximation}~\cite{Kraichnan:PF1965Lagrangian_history}, \textit{Test Field Model}~\cite{Kraichnan:JFM1972}, etc., to derive $k^{-5/3}$ energy spectrum for HDT.  These theories, however, are beyond the scope of this review. We refer a reader to Kraichnan's original papers and \citet{Leslie:book}.

In spite of above warnings by Kraichnan,   Eulerian framework has been successfully employed in a large number of field theory calculations of hydrodynamic, scalars, and MHD turbulence (see Sections~\ref{sec:RG}, \ref{sec:ET}, \ref{sec:PS}, \ref{sec:MHD}).   So, how do we reconcile Kraichnan's objections with the success of Eulerian field theory? Our viewpoint on this topic is as follows. \citet{Verma:INAE2020_sweeping} performed RG analysis of the following equations:
 {\bea
	\frac{\partial {\bf u}}{\partial t} -  \nu \nabla^2 {\bf u} & = &    - {\bf U}_0 \cdot \nabla {\bf u}   - {\bf u} \cdot \nabla {\bf u}  
	- \nabla p + {\bf f},  \label{eq:NS_sweeping}  \\
	\nabla \cdot {\bf u} & = & 0,
	\label{eq:del_u_0_sweeping}
	\eea}	
where ${\bf U}_0 $ is the mean velocity field. The steps of the RG computations remain the same as in  Section~\ref{sec:RG}, except that  the  small frequency limit of Eq.~(\ref{eq:nu_omega_0}) is replaced by the Doppler-shifted frequency $\omega_D = \omega - {\bf U}_0 \cdot {\bf k} \rightarrow 0$. These steps yield $k^{-5/3}$ energy spectrum and the same renormalized viscosity as in Sec.~\ref{sec:RG}.  Thus, \citet{Verma:INAE2020_sweeping}  showed that the renormalized parameters and energy spectrum remain unchanged with ${\bf U}_0$, which is consistent with Galilean invariance. 
However, random fluctuations at large scales do affect the correlations~\cite{He:ARFM2017,Wilczek:PRE2012,Verma:INAE2020_sweeping}, as we describe below.

The \textit{sweeping effect} is apparent in numerical $\bar{C}({\bf k},t-t') $.  Using numerical data of isotropic HDT, it has been shown that
$\bar{C}({\bf k},t-t')$ follows the following form~\cite{Verma:INAE2020_sweeping,He:ARFM2017,Wilczek:PRE2012}:
\be 
\frac{\bar{C}({\bf k},t-t')}{C({\bf k})} =   \exp[-\nu(k) k^2 (t-t')] 
\exp[-i  ck \tilde{U}_0 (t-t') ] \exp[-i {\bf U}_0 \cdot {\bf k} (t-t') ], 
\label{eq:Verama_sweeping}
\ee 
where ${\bf U}_0$ is the mean velocity field, and $ \tilde{U}_0$ is the random large-scale velocity. 
In Eq.~(\ref{eq:Verama_sweeping}), $\exp[-i {\bf U}_0 \cdot {\bf k} (t-t') ]$ is the trivial advection term that is related to Taylor's frozen-in hypothesis~\cite{Taylor:PRS1938}, whereas $\exp[-i  ck \tilde{U}_0 (t-t') ]$ represents sweeping effect by random large-scale velocity.  \citet{Verma:INAE2020_sweeping} argued that $k \tilde{U}_0 \sim k^{2/3}$  will yield $k^{-5/3}$ energy spectrum even in the presence of random sweeping effect; this hypothesis however needs to be tested using high-resolution simulations. In addition, it will be interesting to attempt a RG calculation for $\nu$ and $\tilde{U}_0$ simultaneously.  Thus, sweeping effect remains an enigma even after 60 years of research.

In addition, weak turbulence theory has been applied  to many  other systems, including anisotropic ones. However, we do not discuss this topic any further due to lack of space.  Next, we will briefly discuss field theory of intermittency.

\section{Field Theory for Intermittency}
\label{sec:Intermittency}

In equilibrium field theory, $n$-th order correlation functions and the respective Green's functions are the same, and they have been computed for various systems in the past~\cite{Peskin:book:QFT,Goldenfeld:book,Fradkin:book:QFT}. However, the  Green's function and the  correlation function are different in nonequilibrium field theory, turbulence being one of them [e.g., see Eqs.~(\ref{eq:Gk_tt'}, \ref{eq:Ck_tt'})]. In turbulence field theory,  the second-order and  triple-order correlations of the velocity field have been computed by many researchers.  

\citet{Kolmogorov:DANS1941Dissipation,Kolmogorov:DANS1941Structure} derived the exact 
 third-order structure function   under the assumption of homogeneity and isotropy [see Eq.~(\ref{eq:S3_K41})]. Note  that Eqs.~(\ref{eq:S3_K41}, \ref{eq:fluid_flux}) yield average energy flux $\Pi$.  However, the fluctuations in $\Pi$ have not been computed from the first principle. Similarly,  no one has been able to derive higher-order correlations for \textbf{u} in HDT from the first principles.   

The $q$-th order structure function is defined as~\cite{Frisch:book}
\be 
S_q(l) = \la | ({\bf u(x+l) - u(x)}) \cdot \hat{l} |^q 
\ra  ,
\ee
where $q>0$, often taken as integers from 1 to 10 or so. Phenomenologically, 
\be 
S_q(l) =  \la \Pi_l^{q/3} \ra  l^{q/3} ,
\label{eq:Sq_model}
	\ee
where $\Pi_l$ denotes the fluctuating energy flux at length scale $l$. Note that
\be 
\la \Pi_l \ra = \Pi,~~~\mathrm{but}~~
\la \Pi_l^q \ra \ne \Pi^q
\ee 
because of the nontrivial probability distribution of $ \Pi_l$. 
In literature, the fluctuations in $ \Pi_l$ has been modelled as~\cite{Frisch:book}
\be 
 \la \Pi_l^{q} \ra = A_q \la \Pi \ra^{q} 
 \left( \frac{l}{L} \right)^{\tau_q},
 \label{eq:Pi_l}
 \ee 
where $A_q$'s are constants, and $\tau_q$'s are exponents related to the energy flux. Substitution of Eq.~(\ref{eq:Pi_l}) in Eq.~(\ref{eq:Sq_model}) yields~\cite{Frisch:book}
\be 
S_q(l) = A_{q/3} \la \Pi \ra^{q/3}  l^{q/3+\tau_{q/3}} L^{-\tau_{q/3}}
= A_{q/3} \la \Pi \ra^{q/3}  l^{\zeta_q} L^{-\tau_{q/3}},
\label{eq:Sq_final}
\ee
where
\be 
\zeta_q = q/3+\tau_{q/3}
\ee 
is called the \textit{intermittency exponent}. A simple generalization of Kolmogorov's theory yields  $\zeta_q=q/3$, which is inconsistent with the numerical and experimental observations~\cite{Frisch:book}.  Researchers have constructed various phenomenological models, e.g., \textit{fractal model}~\cite{Frisch:JFM1978}, \textit{lognormal model}~\cite{Kolmogorov:JFM1963}, \textit{multifractal model}~\cite{Meneveau:PRL1987}, \textit{She-Leveque model}~\cite{She:PRL1994}. Among them, She-Leveque model~\cite{She:PRL1994} provides the best fit to the numerical and experimental observations. Note, however, that the above models are not first-principle computations from the NS equation.

In the following, we briefly describe several field-theoretic attempts to compute the intermittency exponents.

\begin{enumerate}
	
\item \citet{Belinicher:JSP1998} developed a field-theoretic procedure to compute  the intermittency exponent $\zeta_q$. In a series of follow-up papers, \citet{Lvov:PRE1995I,Lvov:PRE1995II,Lvov:PRE1996}  derived scaling relations among the intermittency exponents  using exact resummation of all the Feynman diagrams. These computations are divergence-free  in infrared and ultraviolet regimes. The above scaling relations, referred to as \textit{ fusion rules}~\cite{Lvov:PRL1996fusion,Fairhall:PRL1997}, are in  good agreement with the experimental results of atmospheric turbulence.

\item \citet{Onsager:Nouvo1949_SH}, \citet{Frisch:book},  and \citet{Eyink:RMP2006} discussed \textit{Euler singularity} and \textit{dissipative anomaly}. Note that infinitesimal $\nu$  yields a finite energy dissipation or flux, whereas zero viscosity in spectrally-truncated Euler turbulence leads to an equilibrium solution with no energy flux.   \citet{Migdal:PR2023} investigated the inviscid limit of the Navier-Stokes equation and reported anomalous terms in Hamiltonian, dissipation, and helicity.  Migdal and collaborators  also proposed \textit{area law} for HDT. 
These issues have been partially addressed by mathematicians~\cite{Luo:2014PNAS} and field theorists~\cite{Eyink:PRX2018,Migdal:PR2023}, but final word is not yet out.

\item \textit{Functional renormalization} and   \textit{generating functionals} have been employed to compute correlation functions to all orders, both in equilibrium~\cite{Peskin:book:QFT} and nonequilibrium settings~\cite{DeDominicis:PRA1979,Canet:JFM2022,Antonov:RNC2025}. The equilibrium computations have been quite successful, but the  intermittency exponents computed  using nonequilibrium framework differ from the experimental and numerical results~\cite{Canet:JFM2022}. Interestingly, \citet{Moriconi:PRE2004} calculated the statistics of intense turbulent vorticity events using Martin-Siggia-Rose functional framework; these results are in good agreement with numerical simulations.  These computations  are quite complex, and they are beyond the scope of this paper.

\item \citet{Das:EPL1994} employed mode coupling method to compute the second-order correlation for the energy flux, $\la \Pi({\bf x})\Pi({\bf x+l}) \ra$.  A generalization of the above computation to higher orders may yield statistics on the fluctuations in $\Pi_l$, which is a crucial factor in  modelling intermittency.

\item Intermittency exponents have been computed exactly for the  passive scalar turbulence with the advective velocity field correlated in space, but delta-correlated in time (given by  Kraichnan model)~\cite{Falkovich:RMP2001}.  These calculations, based on semi-Lagrangian field-theoretic approach, are quite complex. A reader may refer to the original papers cited in the review article~\cite{Falkovich:RMP2001}. There have been attempts to generalize these calculations to HDT, but they have  been unsuccessful so far.

\end{enumerate}

In the next section, I compare turbulence field theory with other field theories.

\section{Comparing Turbulence with Other Field Theories}
\label{sec:other_FT}

From a dynamical perspective, field theories are broadly classified into the following three categories. 
\begin{enumerate}
	\item \textit{Equilibrium field theories} describe systems in  equilibrium with heat bath (in classical field theory) or with quantum noise (in quantum field theory).   Examples of such field theories include Ising model, Wilson's $\phi^4$ theory, and Hubbard model. Here, the fields are Gaussian or quasi-Gaussian, and they respect detailed balance.  The Green's and correlation functions are the same in these field theories.
	
	It is conjectured that  isolated multi-particle  systems (both classical and quantum) thermalize asymptotically~\cite{Srednicki:PRE1994}.  In turbulence, spectrally-truncated Euler equation $(\nu=0 )$ and Gross-Pitaevskii (GP) equation with zero potential are isolated systems that thermalize asymptotically~\cite{Cichowlas:PRL2005,Barenghi:arxiv2016,Nazarenko:book:WT}. Typically, these systems exhibit equipartition of energy or particles among the available Fourier modes,  and zero energy or particle flux (see Sections~\ref{sec:RG} and \ref{sec:ET}).  Researchers have invoked several  mechanisms, including \textit{Berry's conjecture}~\cite{Berry:book_chapter:2020}, to explain thermalization. In turbulence,  the forward energy flux from large scales to small scales plays a key role in the redistribution of energy towards thermalization.  \citet{Verma:EPJB2019,Verma:EPL2025} has argued that the energy flux may work as a thermalization mechanism in isolated multi-particle systems.
	
	It is known that several quantum systems do not thermalize, with prime examples being   \textit{many body localization}~\cite{Abanin:RMP2019} and \textit{time crystals}~\cite{Sacha:ROPP2017}.   For many body localization, emergent integrability and multiple conserved quantities are believed to be the key factors for nonthermalization~\cite{Abanin:RMP2019}. Interestingly, 2D Euler turbulence does not thermalize for some ordered initial condition, possibly due  to its multiple conserved quantities~\cite{Verma:PRF2022}.  Thus,  2D Euler turbulence and many body localization appear to share several common features.  This issue needs further exploration.

	\item \textit{Near-equilibrium systems} are out of equilibrium, but they are quite close to equilibrium.  A popular example is  thermal conduction where the directional heat transport breaks detailed balance. Systems near equilibrium obey \textit{fluctuation-dissipation theorem}~\cite{Pathria:book}.  To illustrate, for the diffusion equation,
	\be 
	\partial_t \phi({\bf k},t) = -\kappa k^2 \phi({\bf k},t) + f_\phi({\bf k},t),
	\ee 
	where $f_\phi$ is the white noise, the Green's function is
	\be 
	G({\bf k}, t-t') = H(t-t') \exp[-\kappa k^2 (t-t')],
	\label{eq:Green_FD}
	\ee
	with $H(t-t')$ as the step function. For this system, the	fluctuation-dissipation theorem yields
	\be 
	|\phi({\bf k},\omega)|^2 = C({\bf k}, \omega) 
	= \frac{k_B T}{\omega} \Im [G({\bf k}, \omega) ],
	\ee
	or
	\be 
	C({\bf k}, t-t') = C({\bf k}) \exp[-\kappa k^2 (t-t')],
	\label{eq:C_kappa_FD}
	\ee
	where 
	\be C({\bf k})  = |f_\phi({\bf k},t)|^2 = k_B T
	\ee
	is the equipartitioned thermal energy of each Fourier mode. Note that the fluctuation-dissipation theorem tells us how the correlation function $	C({\bf k}, t-t')$ and Green's function $G({\bf k}, t-t') $  decay with time.

	\item \textit{Nonequilibrium field theories}  describe systems that are far from equilibrium. A typical nonequilibrium system has  an external driving at large scales and dissipation at small scales. Some prominent examples are turbulence~\cite{Lesieur:book:Turbulence}, \textit{active matter}~\cite{Alert:ARCMP2021}, \textit{coarsening systems}~\cite{Hohenberg:RMP1977,		Verma:PRE2023_coarsening}, and \textit{driven quantum turbulence} (e.g.,  \textit{superfluids})~\cite{Madeira:ARCMP2020}.  Some systems, e.g.,  \textit{Kardar-Parisi-Zhang} (\textit{KPZ}) equation~\cite{Kardar:PRL1986}, are forced by white noise. 	Nonequilibrium systems exhibit directional energy transfers, e.g., forward energy transfers in 3D turbulence. In addition, nonequilibrium fields are non-Gaussian and time-dependent; and the corresponding Green's function and correlation function are frequency dependent, and they are not the same.
	
	Field-theoretic frameworks for equilibrium and  nonequilibrium systems have similarities. Most calculations attempt to derive RG flow equations for the diffusive parameter, coupling constant, and forcing amplitude (see \cite{Kardar:PRL1986,Yakhot:JSC1986,Forster:PRA1977}).  In HDT,   the renormalized viscosity has been computed using   Yakhot-Orszag's formalism, recursive RG, and functional RG.  Note, however, that the coupling constant of HDT remains unchanged under RG due to the Galilean symmetry, whereas the assumption of large-scaling forcing keeps the  forcing amplitude  unnormalized. The energy flux  for HDT too has been computed using perturbative field theory.  Apart from turbulence, there are only a handful of field theory works on the energy transfers, some of which are  \citet{Bratanov:PNAS2015} for active turbulence, and  \citet{Verma:PRE2023_coarsening} and \citet{Yadav:PRE2024} for coarsening systems. 
	 
	For nonequilibrium systems, researchers employ a generalized  fluctuation-dissipation theorem. One instance of such an attempt is Eqs.~(\ref{eq:Gk_tt'}, \ref{eq:Ck_tt'}), where the diffusion coefficient of  Eqs.~(\ref{eq:Green_FD}, \ref{eq:C_kappa_FD}) is replaced by its renormalized counterpart, $\nu(k)$. In addition,  $C({\bf k})$ of Eq.~(\ref{eq:Ck_tt'}) is nonthermal, and it is derived using other means, e.g., dimensional analysis for HDT~\cite{Leslie:book}.  Note that Eqs.~(\ref{eq:Gk_tt'}, \ref{eq:Ck_tt'}) are phenomenological, and they have not been derived from the first principle.
	
\end{enumerate}

\begin{figure}
	\begin{center}
		\includegraphics[scale = 1]{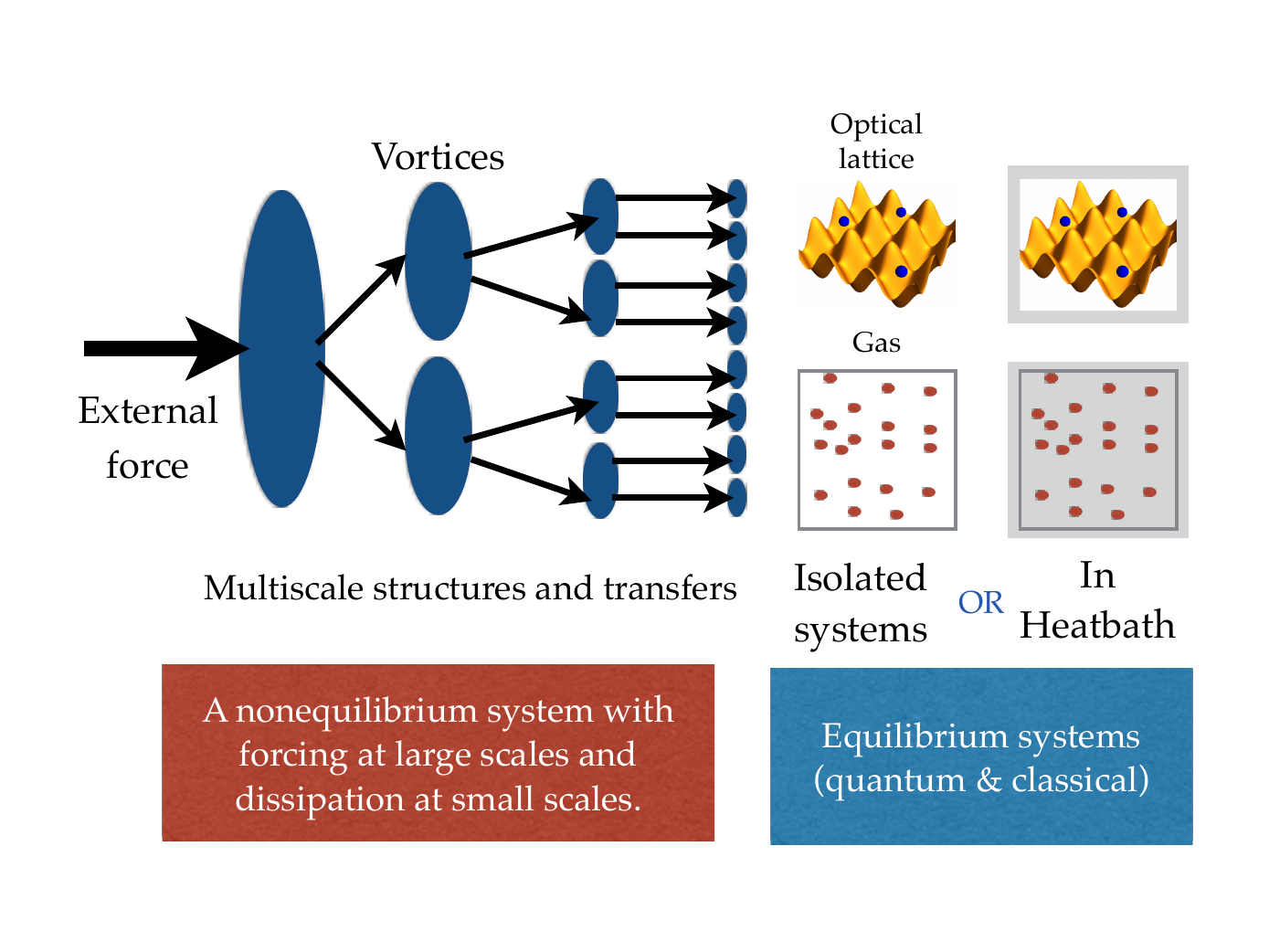}
	\end{center}
	\vspace*{0pt}
	\caption{Schematic diagrams of equilibrium and nonequilibrium systems. Equilibrium systems are either isolated or in contact with heat bath (shown as shaded region), whereas the nonequilibrium ones are typically forced at large scales~\cite{Verma:EPJB2019}. Picture of optical lattice taken from wikipedia.}
	\label{fig:equil_nonequil}
\end{figure}

The above connections among multidisciplinary field theoretic works  are encouraging.  Note that many field-theoretic computations of driven dissipative systems employ  partial differential equations that are typically forced and dissipative.  Thus,   Hamiltonian or Lagrangian are not mandatory for field theory applications, contrary to a popular belief.

In Fig.~\ref{fig:equil_nonequil} we contrast nonequilibrium systems with equilibrium ones.  Nonequilibrium systems are typically multiscale with forcing at large scales and dissipation at small scales; this scale separation between the forcing and dissipation yields energy cascade from large scales to small scales\footnote{In some systems, multiple conservation laws lead to an inverse energy cascade. Two-dimensional hydrodynamics is one such example.}. In contrast, equilibrium systems are either isolated or in contact with heat bath, and they typically have  a single scale (e.g., mean free path length in an ideal gas).  In addition, equilibrium systems respect detailed balance and have zero energy flux, whereas nonequilibrium ones break detailed balance and exhibit nonzero energy flux.

There are several exceptions to the above classification.  Three-dimensional Euler turbulence with ordered initial condition exhibits forward energy flux at large scales, but zero flux at small scales. Hence, 3D Euler turbulence is out of equilibrium at large scales and in equilibrium at small scales~\cite{Cichowlas:PRL2005,Verma:PRF2022,Verma:EPL2025}.  Similarly, atmospheric flows exhibit nonequilibrium  behaviour at large and intermediate scales, but near-equilibrium behaviour at small scales. This is the reason why we can define local temperature for the atmosphere even when the large scales are highly dynamic.

I conclude the review in the next section.

\section{Discussions and Summary}

First principle calculations are rare in turbulence. Fortunately, in addition to the exact results by Kolmogorov, there are several  analytical results based on field theory. In this review, I cover past works on viscosity  renormalization and energy flux computations of turbulence.   Starting with Kraichnan's direct interaction approximation, I discuss Yakhot-Orszag's renormalization group (RG) formalism, recursive RG by McComb and Zhou,  functional RG, and RG scheme using Craya-Herring basis.  In particular, the limitations and benefits of these methods are compared. For example,   infrared divergence in direct interaction approximation is cured in all the RG schemes. The recursive RG scheme of McComb and Zhou is self-consistent with $k^{-5/3}$ energy spectrum, whereas the energy spectrum in Yakhot-Orszag's scheme depends on the forcing function.  Another  kind of perturbative field theory calculations yield the energy flux and  Kolmogorov's constant for HDT. The results from these computations are consistent with numerical and experimental observations.  

Note that the spectrally-truncated  Euler equation $(\nu=0 )$ is an isolated system that yields a very different solution. In thermalized Euler turbulence, the available Fourier modes have equal energy that yields zero energy flux~\cite{Cichowlas:PRL2005}. 	However, \citet{Verma:PRF2022} showed that 2D Euler turbulence does not thermalize for some ordered initial conditions.

The field-theoretic calculations of HDT have been extended to other flows, e.g., passive scalar turbulence and MHD turbulence. Note, however, that field-theoretic computation of anisotropic turbulence still remains a challenge~\cite{Sagaut:book,Alexakis:PR2018}. For example, we do not have successful   field-theoretic calculations for stably stratified turbulence, turbulent convection,  rotating turbulence, and anisotropic MHD turbulence (in the presence of a mean magnetic field).  Interestingly, computation of anisotropic  energy flux  in weak turbulence framework is reasonably well developed~\cite{Zakharov:book:WaveTurb,Nazarenko:book:WT}. 

Field theory of turbulence share similarities with other  nonequilibrium field theories, including Kardar-Parisi-Zhang (KPZ) equation~\cite{Kardar:PRL1986},  coarsening systems~\cite{Hohenberg:RMP1977}, and active turbulence~\cite{Bratanov:PNAS2015}.  All these computations invoke external noise, which may be thermal or nonthermal.  I believe that a detailed comparison between these computations would yield interesting insights.

Phenomenologies of compressible and incompressible turbulence have significant differences. For example, shock-dominated compressible flows exhibit $k^{-2}$ energy spectrum, rather than $k^{-5/3}$ spectrum. Interestingly, superfluid turbulence has a strong connection with compressible turbulence. Unfortunately, there are only a limited number of field-theoretic works on compressible and superfluid turbulence~\cite{Schmitt:LNP2015}.  It is hoped that such computations would be performed in future.

\vspace{0.5cm}

{\bf Acknowledgements}: The author thanks Srinivas Raghu, Pankaj Mishra, and J. K. Bhattacharjee for useful discussions. He
received valuable ideas during the discussion meeting \textit{Field Theory
and Turbulence} hosted by International Centre for Theoretical Studies, Bengaluru.
This work is supported by J. C. Bose Fellowship (Grant No. SERB/PHY/2023488), and by the Science and Engineering Research Board, India (Grant
numbers: SERB/PHY/20215225 and SERB/PHY/2021473).




\end{document}